\pdfoutput=1 
\documentclass[12pt]{article}

\usepackage{graphicx}
\usepackage{xcolor}
\usepackage[unicode,breaklinks=true,linktocpage=true]{hyperref} 
\usepackage{ascmac}
\usepackage{amsmath,amssymb,amsthm}
\usepackage{amscd} 
\usepackage[all]{xy}
\usepackage{mathrsfs} 
\numberwithin{equation}{section} 
\usepackage[top=0.14\paperwidth,bottom=0.14\paperwidth,left=0.14\paperwidth,right=0.14\paperwidth]{geometry}

\usepackage{caption}
\usepackage{cite}
\DeclareMathOperator{\im}{Im}
\DeclareMathOperator{\diag}{diag}
\DeclareMathOperator{\link}{Link}

\DeclareMathOperator{\rank}{rank}

\newcommand{\nn}{\nonumber}
\newcommand{\ig}{\includegraphics}
\newcommand{\weds}{\wedge \cdots \wedge }

\newcommand{\corr}{\leftrightarrow}

\newcommand{\vevs}[1]{\langle #1 \rangle}

\newcommand{\er}[1]{eq.~\eqref{#1}}
\newcommand{\ers}[1]{eqs.~\eqref{#1}}
\newcommand{\qtq}[1]{\quad\text{#1}\quad}

\newcommand{\bb}{\mathbb}
\newcommand{\na}{\nabla}
\newcommand{\hph}{\hphantom}
\newcommand{\ri}{\mathrm{i}}
\newcommand{\rd}{\mathrm{d}}
\newcommand{\re}{\mathrm{e}}

\renewcommand{\b}{\bar}

\newcommand{\sr}{\sqrt}
\newcommand{\bs}{\boldsymbol}

\newcommand{\fr}{\frac}

\newcommand{\der}{\partial}

\renewcommand{\(}{\left(}
\renewcommand{\)}{\right)}

\newcommand{\wed}{\wedge}
\newcommand{\bmx}{\left(\begin{matrix}}
\newcommand{\emx}{\end{matrix}\right)}
\newcommand{\mtx}[1]{\bmx #1 \emx}

\begin{document}

\allowdisplaybreaks[4]
\begin{titlepage}
\renewcommand{\thefootnote}{\fnsymbol{footnote}}%
 \vspace{3em}
 \begin{center}%
  {\LARGE 
Generalized chiral instabilities, linking numbers, \\
and non-invertible symmetries
  \par
   }
 \vspace{1.5em}
  { \large
Naoki Yamamoto\footnote{nyama@rk.phys.keio.ac.jp}${}^a$ and 
Ryo Yokokura\footnote{ryokokur@keio.jp}${}^{b}$
   \par
   }
  \vspace{1em} 
${}^a${\small\it Department of Physics, Keio University, 
Hiyoshi 3-14-1, Yokohama 223-8522, Japan}
\par
${}^b${\small\it 
Department of Physics \& 
Research and Education Center for Natural Sciences,
\par 
Keio University, Hiyoshi 4-1-1, Yokohama, 223-8521, Japan}
 \end{center}%
\vspace{1.5em}

\begin{abstract}
We demonstrate a universal mechanism of a class of instabilities in infrared regions for massless Abelian $p$-form gauge theories with topological interactions, which we call generalized chiral instabilities. Such instabilities occur in the presence of initial electric fields for the $p$-form gauge fields. We show that the dynamically generated magnetic fields tend to decrease the initial electric fields and result in configurations with linking numbers, which can be characterized by non-invertible global symmetries. The so-called chiral plasma instability and instabilities of the axion electrodynamics and $(4+1)$-dimensional Maxwell-Chern-Simons theory in electric fields can be described by the generalized chiral instabilities in a unified manner. We also illustrate this mechanism in the $(2+1)$-dimensional Goldstone-Maxwell model in electric field.
\end{abstract}
\end{titlepage}
 \setcounter{footnote}{0}%
\renewcommand{\thefootnote}{$*$\arabic{footnote}}%
\tableofcontents

\newpage

\section{Introduction and summary}
Electromagnetism exhibits instabilities in the presence of certain background fields and/or dynamical fields. Examples include the chiral instabilities in $(3+1)$-dimensional electromagnetism with Lorentz- and parity-violating background field~\cite{Carroll:1989vb}, such as a time-dependent background axion field~\cite{Anber:2006xt}, or with fermion chirality imbalance~\cite{Joyce:1997uy,Akamatsu:2013pjd}%
\footnote{While instabilities in the presence of chirality imbalance in media are called chiral plasma instabilities~\cite{Akamatsu:2013pjd}, those in the parity-breaking background field in the vacuum will be referred to simply as chiral instabilities in this paper.}
and the instabilities in the axion electrodynamics~\cite{Bergman:2011rf,Ooguri:2011aa} and $(4+1)$-dimensional Maxwell-Chern-Simons theory~\cite{Nakamura:2009tf} in electric fields. These instabilities have also been applied to various systems, such as the early Universe, neutron stars~(see ref.~\cite{Kamada:2022nyt} for a review), and topological materials~\cite{Ooguri:2011aa}.

In this paper, we argue that the all above examples can be universally described as instabilities in Abelian gauge theories for massless $U(1)$ $p$-form gauge fields ($p = 0,1,2,...$) with topological couplings that are second order in derivatives and third order in gauge fields. We will call them the generalized chiral instabilities. 
We consider these theories in $D$-dimensional flat Minkowski spacetime under the initial configurations of homogeneous ($p$+1)-form electric fields for some of the gauge fields.

Based on the linear order analysis, we show that the gauge fields that are coupled to the initial electric fields become tachyonic in infrared regions, leading to the generation of magnetic fields. Furthermore, the dynamics of the tachyonic modes decreases the initial electric fields so that the instabilities tend to be weakened. We also find that the produced magnetic fields have linked configurations characterized by a generalization of the magnetic helicity widely applied in plasma physics and magnetohydrodynamics (see, e.g., ref.~\cite{Davidson2001}). Similarly to the conventional magnetic helicity, however, this generalized magnetic helicity is not invariant under large gauge transformations. We develop its gauge-invariant formulation in terms of the non-invertible global symmetries (see ref.~\cite{Cordova:2022ruw} for a review), which allows one to rephrase the chiral anomaly~\cite{Adler:1969gk,Bell:1969ts} as certain symmetries at the expense of invertibility~\cite{Choi:2022jqy,Cordova:2022ieu,Choi:2022fgx,Yokokura:2022alv,Putrov:2023jqi} (see also refs.~\cite{GarciaEtxebarria:2022jky,Karasik:2022kkq}).
We confirm the relation between the generalized magnetic helicities and the linking numbers in terms of the non-invertible symmetries.

We note that the number of these tachyonic modes can be counted similarly to the usual Nambu-Goldstone modes if we regard the Abelian gauge fields as Nambu-Goldstone modes of higher-form global symmetries~\cite{Gaiotto:2014kfa} in certain background fields~\cite{Yamamoto:2022vrh}. In this paper, on the other hand, we do not assume whether gauge fields can be seen as Nambu-Goldstone modes or not, such that we can apply the generalized chiral instabilities in broader physical situations.

There are several avenues for future work. An immediate question is the final field configuration after the generalized chiral instabilities. To answer this, one needs to perform a full non-linear analysis. It would be particularly interesting to see if the turbulent behaviors are modified in the presence of the generalized magnetic helicity. In the case of the conventional chiral instability, the non-linear evolution displays the inverse cascade, leading to the large-scale structure of magnetic fields; see, e.g., ref.~\cite{Kamada:2022nyt}. Incorporating the effects of the gravity would also be interesting, e.g., in light of its possible applications to the string theory and gauge/gravity duality. While we assume the flat Minkowski spacetime in this paper, it is known in the $(4+1)$-dimensional Maxwell-Chern-Simons theory that the presence of the gravity (such as the AdS background) can eliminate the instability~\cite{Nakamura:2009tf}.

This paper is organized as follows.
In section~\ref{CPI}, we review the conventional chiral instability including the presence of the unstable modes, reduction of the initial field, and relation between the magnetic helicity and the linking number. We also give a new understanding of the magnetic helicity from the viewpoint of the non-invertible symmetries.
In section~\ref{GCI}, we generalize the above chiral instability to the instabilities for $p$-form Abelian gauge theories with the topological couplings in $D$ dimensions. We show the reduction of the initial electric field and the generation of generalized magnetic helicities. We give a gauge-invariant formulation of the generalized magnetic helicities based on non-invertible symmetries. 
In section~\ref{appli}, we apply the mechanism of the generalized chiral instabilities to the axion electrodynamics and $(2+1)$-dimensional Goldstone-Maxwell model in the background electric field. We clarify the physical meanings of the generalized magnetic helicities and their relations to linking numbers in both cases.

We use the conventions for the spacetime metric $\eta_{\mu\nu} = \diag (-1, +1,...,+1)$, and the totally anti-symmetric tensor $\epsilon_{01...D-1} = +1$. The indices with the Greek letters denote the spacetime directions $\mu,\nu,... = 0,1,..., D-1$, and those of the Roman letters denote the spatial directions $i,j,...= 1,2,..., D-1$.

\section{\label{CPI}Chiral instability}

In this section, we review the chiral instability in $(3+1)$-dimensional spacetime. Although most of the results in this section are known in the literature, we will reformulate them through different paths so that they can be extended to a more generic case with $p$-form gauge fields in section~\ref{GCI}. We will also show the relationship between the magnetic helicity generated via the chiral instability and the linking number from the viewpoint of the non-invertible symmetry.

\subsection{\label{CPImode}Unstable modes}
We first show the presence of one unstable mode in the theory described by the following action for a $U(1)$ 1-form gauge field $a = a_\mu {\rd}x^\mu$ and an axion field $\phi$:
\begin{equation}
 S = - \int {\rd}^4 x \(
 \fr{1}{4e^2} f_{\mu \nu} f^{\mu\nu}
+
\fr{v^2}{2} 
\der_\mu \phi \der^\mu \phi
+ \fr{1}{16 \pi^2} \phi 
f_{\mu\nu } \tilde{f}^{\mu \nu}
\),
\label{CPIaction}
\end{equation}
where $f_{\mu\nu} = \der_\mu a_\nu - \der_\nu a_\mu$ and $\tilde{f}^{\mu \nu} = \fr{1}{2}\epsilon^{\mu\nu\rho \sigma} f_{\rho\sigma}$. Here, we can regard $\phi$ as a $U(1)$ 0-form gauge field that has the redundancy under the shift $\phi \to \phi + 2\pi$~\cite{Cordova:2019jnf,Cordova:2019uob}. 
The equations of motion for $\phi$ and $a$ are 
\begin{gather}
v^2 \der_\mu \der^\mu \phi
- \fr{1}{16\pi^2} f_{\mu\nu} \tilde{f}^{\mu\nu} 
=0\,,
 \label{EOM_phi}
 \\
 \fr{1}{e^2}\der_\mu f^{\mu\nu} +
\fr{1}{8\pi^2}\epsilon^{\mu\nu \rho\sigma} (\der_\mu \phi ) f_{\rho\sigma}
=0\,,
\label{EOM_a}
\end{gather}
respectively. We assume that $\phi$ has the initial value $\b\phi$ with a homogeneous configuration $\der_0 \b\phi \neq 0$ and $\der_i \bar\phi =0$.

The spatial ($\nu=i$) component of \er{EOM_a} is the modified Maxwell-Amep\`ere law,
\begin{equation}
\bs{\na} \times \bs{B} = \bs{j} + \der_0 \bs{E}, \qquad 
\bs{j} = \b\sigma \bs{B},
\label{CME}
\end{equation}
where $E_i = -f_{0i}$ and $B_i=\tilde{f}_{0i}$ denote the electric and magnetic fields, respectively, and $\bar \sigma = C(\der_0 \b\phi)$ with $C=\frac{e^2}{4\pi^2}$. The electric current proportional to the magnetic field in \er{CME} is the so-called chiral magnetic effect~\cite{Vilenkin:1980fu,Nielsen:1983rb,Fukushima:2008xe} if $\der_0 \b\phi$ is identified as $\der_0 \b\phi = 2\mu_5$ with $\mu_5$ being a chiral chemical potential. This $\der_0 \bar\phi$ can also be regarded as an initial electric field for the $0$-form gauge field $\phi$ in a way similar to the usual electric field $\bar{E}_i = - \der_0 \bar{a}_i$ for a 1-form gauge field $a$ after gauge fixing. In section~\ref{GCI}, we will generalize it to the electric field for $p$-form gauge fields.

To see the presence of the unstable mode, we rewrite the equation of motion in \er{EOM_a} in momentum space as
\begin{equation}
     (\omega^2 - |\bs{k}|^2) a^i 
    = \b\sigma \epsilon^{i l j 0 } 
    \ri k_l a_j  \,,
\label{EOM_a_momentum}
\end{equation}
where we have taken the temporal gauge $a_0 =0$ with the Gauss law constraint $k_i a^i =0$.
The dispersion relation then reads
\begin{equation}
\label{dispersion_CI}
    \omega^2 = |\bs{k}|^2 \pm |\b\sigma \bs{k}|\,.
\end{equation}
Therefore, we have one unstable mode in the infrared region, $ 0 < |\bs{k}| < |\b\sigma|$. At the critical wave number $|\bs{k}| = |\b\sigma|$, we have a static mode satisfying the relation $\bs{\na} \times \bs{B} = \b\sigma \bs{B}$. This relation can also be written as ${\na}^2 a_i = - \b\sigma \epsilon_{ijk} \der_j a_k$ in terms of the gauge field, with $\epsilon_{ijk}$ being the anti-symmetric tensor with $\epsilon^{123} = \epsilon_{123} = +1$. This field is the so-called Beltrami field (see, e.g., ref.~\cite{Davidson2001}).

\subsection{\label{CPIreduction}Reduction of the initial field}

We next show that the unstable mode tends to reduce the initial $|\der_0 \b\phi|$ at the level of linear analysis. To see this, we decompose $\phi$ as $\phi = \b\phi + \delta \phi$, where $\bar{\phi}$ is the initial configuration and $\delta \phi$ is the fluctuation.

Let us consider the time evolution of $\der_0 \phi$. By multiplying the equation of motion in \er{EOM_phi} by $\der_0 \bar \phi$ and integrating over space, we find
\begin{align}
v^2 \int {\rd}^3 \bs{x}
\der_0 \bar\phi \der_0 (\der_0 \delta \phi )
&= 
- \fr{1}{16\pi^2} \int {\rd}^3 \bs{x}
\der_0 \bar\phi f_{\mu\nu} \tilde{f}^{\mu\nu} 
= \frac{1}{4\pi^2}  \int {\rd}^3 \bs{x}
\der_0 \bar{\phi} \epsilon^{ i l j 0 } \der_0 a_i  \der_l a_j \,. 
\label{mu5_evolution1}
\end{align}
We will show that \er{mu5_evolution1} is negative, so the sign of $\der_0 (\der_0 \delta \phi)$ is opposite to that of $\der_0 \bar\phi$, implying the reduction of the initial field.

To see how the time evolution of $\der_0 \phi$ is related to the unstable modes, we perform the Fourier decomposition of the normalized gauge field $\hat{a}_i = \fr{1}{e} a_i$ in the temporal gauge $a_0 =0$,
\begin{equation}
 \hat{a}_i (x)  =
 \int \fr{{\rd}^3 \bs{k}}{(2\pi)^3} 
\hat{a}_i (t, \bs{k}) {\re}^{{\ri}\bs{k} \cdot \bs{x}}.
\end{equation}
For $\hat{a}_i(x)$ to be real, $\hat{a}_i (t,\bs{k})$ satisfies the condition $\hat{a}_i^*(t,\bs{k}) = \hat{a}_i (t,-\bs{k})$.
We can then rewrite \er{mu5_evolution1} as
\begin{align}
 v^2 \int {\rd}^3 \bs{x}
\der_0 \bar\phi \der_0 (\der_0 \delta \phi )
= 
\int \fr{{\rd}^{3} \bs{k}}{(2\pi)^{3}} 
\der_0 \hat{a}_i (t, -\bs{k}) \ri \b\sigma \epsilon^{i l j 0 } k_l \hat{a}_j (t, \bs{k})\,.
\label{mu5_evolution2}
\end{align}
As $\epsilon^{ilj0}$ is anti-symmetric under the  exchange of $i \corr j$, we can block diagonalize the matrix $\b\sigma k_l \epsilon^{ilj0}$ by an orthogonal matrix $P_{ij}$ as
\begin{equation}
 P_{ik} \b\sigma \epsilon^{ k l m 0} k_l P_{m j} = 
\Lambda_{ij},
\label{P_CI}
\end{equation}
where $\Lambda_{ij}$ is a block diagonalized matrix
\begin{equation}
\Lambda_{ij}
:= 
 \mtx{0 & \lambda 
\\
- \lambda  & 0 
\\
&& 0
}
,\quad
 \lambda  = |\b\sigma \bs{k}|.
\end{equation}

We can also block diagonalize the equation of motion in \er{EOM_a_momentum} by the same matrix $P_{ij}$ as
\begin{equation}
(-\der_0^2 - \bs{k}^2) v_i (t,\bs{k}) 
=  \ri \Lambda_{ij} v_j  (t,\bs{k}) ,
\end{equation}
where $v_i$ is a vector satisfying $\hat{a}_i  = P_{ij} v_j$. This equation can be solved as
\begin{align}
 v_j (t,\bs{k}) 
&
 =  
\epsilon^{\rm L}_j
(A^{+}_{\bs{k},\lambda} {\re}^{{\ri} t  \sr{|\bs{k}|^2 +\lambda} }
 + A^{-}_{\bs{k},\lambda} 
{\re}^{-{\ri} t \sr{|\bs{k}|^2 +\lambda}})
\nonumber \\
&\quad
+
\epsilon^{\rm R}_j
(A^{+}_{\bs{k},-\lambda} {\re}^{{\ri} t  \sr{|\bs{k}|^2 -\lambda} }
 + A^{-}_{\bs{k},-\lambda}
 {\re}^{-{\ri} t \sr{|\bs{k}|^2 -\lambda}})
+\cdots
\end{align}
for $|\bs{k}|^2 > \lambda$,
and 
\begin{align}
 v_j (t,\bs{k}) 
&
 =
\epsilon^{\rm L}_j
(A^{+}_{\bs{k},\lambda} {\re}^{{\ri} t  \sr{|\bs{k}|^2 +\lambda} }
 + A^{-}_{\bs{k},\lambda} 
{\re}^{-{\ri} t \sr{|\bs{k}|^2 +\lambda}})
\nonumber \\
&\quad
+
\epsilon^{\rm R}_j
(A^{+}_{\bs{k},-\lambda} \re^{ t\sr{\lambda - |\bs{k}|^2} }
 + A^{-}_{\bs{k},-\lambda} 
 \re^{- t \sr{\lambda  - |\bs{k}|^2}}
)
+\cdots
\end{align}
for $|\bs{k}|^2 < \lambda$. Here, $\epsilon^{h}_j = \fr{1}{\sr{2}}({1, \mp \ri},0)$ with $h = {\rm L,R}$ are polarization vectors, and $A^{\pm}_{\bs{k}, \pm \lambda}$ are the amplitudes for given $\bs{k}$, $\pm \lambda$, and the backward ($+$) and forward ($-$) directions. In the above solutions, we have not included the mode associated with the zero eigenvalue of $\Lambda_{ij}$, which is a longitudinal mode and does not contribute to the right-hand side of \er{mu5_evolution2}. Note that the amplitudes $A^{\pm}_{\bs{k}, \pm \lambda}$ are subject to the reality conditions, 
$(A^{\pm}_{\bs{k}, \lambda})^*
= A^{\mp}_{-\bs{k}, \lambda}$, 
$(A^{\pm}_{\bs{k}, -\lambda})^* 
= 
A^{\mp}_{-\bs{k}, -\lambda}$
for $|\bs{k}|^2 > \lambda$, and
$(A^{\pm}_{\bs{k}, -\lambda})^* 
= A^{\pm}_{-\bs{k}, -\lambda}$
for $|\bs{k}|^2 < \lambda$. We summarize the derivation in appendix~\ref{realCPI}.

In terms of $v_j (t, \bs{k})$, \er{mu5_evolution2} can be written as
\begin{equation}
 v^2 \int {\rd}^3 \bs{x}
\der_0 \bar\phi \der_0 (\der_0 \delta \phi )
= 
\int \fr{{\rd}^3 \bs{k}}{(2\pi)^3}
\der_0 v_i^* (t, \bs{k}) \ri \Lambda_{ij}
v_j  (t,\bs{k})\,.
\label{mu5_evolution3}
\end{equation}
By substituting the solutions with the reality conditions and using $\epsilon_i ^{\rm R} \ri \Lambda_{ij} \epsilon^{\rm L}_j
= - \epsilon_i ^{\rm L} \ri \Lambda_{ij} \epsilon^{\rm R}_j = \lambda$, and $\epsilon_i^{h} \ri \Lambda_{ij} \epsilon^{h}_j = 0$, we can decompose it into the contributions of the positive- and negative-eigenvalue ($\pm\lambda$) modes:
\begin{align}
&
\int \fr{{\rd}^3 \bs{k}}{(2\pi)^3}
\der_0 v_i^* (t, \bs{k}) \ri \Lambda_{ij}
v_j  (t,\bs{k})
\nonumber \\
&
=
- 
\int_{|\bs{k}|^2 < \lambda} 
\fr{{\rd}^3 \bs{k}}{(2\pi)^3}
 \lambda\sr{ \lambda - |\bs{k}|^2 }
(|A^{+}_{\bs{k}, -\lambda}|^2
{\re}^{2 t \sr{\lambda - |\bs{k}|^2 }}
- 
|A^{-}_{\bs{k}, -\lambda}|^2
{\re}^{-2 t \sr{\lambda - |\bs{k}|^2 }}
)
\nonumber \\
&
\quad
-
2 
\int_{|\bs{k}|^2 > \lambda}  \fr{{\rd}^3 \bs{k}}{(2\pi)^3}
\lambda \sr{|\bs{k}|^2 - \lambda}
\im \left[
(A^{-}_{\bs{k}, -\lambda})^*
 A^{+}_{\bs{k}, -\lambda} 
{\re}^{2{\ri}  t \sr{|\bs{k}|^2 - \lambda}}
\right]
\nonumber \\
&
\quad
-
2 
\int \fr{{\rd}^3 \bs{k}}{(2\pi)^3}
\lambda \sr{|\bs{k}|^2 + \lambda}
\im \left[
(A^{-}_{\bs{k}, \lambda})^*
 A^{+}_{\bs{k}, \lambda} 
{\re}^{2{\ri}  t \sr{|\bs{k}|^2 + \lambda}}
\right]\,.
\end{align}
Here, we used
\begin{align}
\int \fr{{\rd}^3 \bs{k}}{(2\pi)^3} 
\lambda 
\sr{|\bs{k}|^2 + \lambda } 
|A^{+}_{\bs{k}, \lambda}|^2 
&
= 
\int \fr{{\rd}^3\bs{k}}{(2\pi)^3} 
\lambda 
\sr{|\bs{k}|^2 + \lambda } 
|A^{-}_{\bs{k}, \lambda}|^2\,,
\end{align}
which can be obtained by the reality condition. In the presence of non-zero amplitude of the unstable modes, $|A^{+}_{\bs{k}, -\lambda}|^2$, the right-hand side becomes negative under the time evolution. Combined with \er{mu5_evolution3}, we arrive at
\begin{equation}
 v^2 \int {\rd}^3 \bs{x}
\der_0 \bar\phi \der_0 (\der_0 \delta \phi )
< 0.
\end{equation}
Therefore, the spatial average of $\der_0 \phi$ decreases, which tends to relax the instability.

\subsection{\label{CPImagheli}Generation of linked magnetic fields}

Let us show that a magnetic field with a linked structure is generated as a consequence of the chiral instability. Here and below, we use differential forms in addition to the tensorial notation in order to simplify the notation. 

Since the equations of motion in \ers{EOM_phi} and \eqref{EOM_a} are written as total derivatives, we have the corresponding conserved quantities~\cite{Sogabe:2019gif,Hidaka:2020iaz,Hidaka:2020izy}:
\begin{align}
Q_\phi ({\cal V})
&
= \int_{\cal V} \(v^2 *\rd \phi + \fr{1}{8\pi^2} a \wed \rd a\)\,,
\label{Q_phi}
\\
Q_a ({\cal S}) 
&
=
\int_{\cal S} 
\( - \fr{1}{e^2} * \rd a + \fr{1}{4\pi^2} \phi  \rd a\)\,,
\label{Q_a}
\end{align}
where ${\cal V}$ and ${\cal S}$ are closed 3- and 2-dimensional subspaces,
respectively. 

By taking ${\cal V}$ as a time slice $V$, we have 
\begin{equation}
 Q_\phi (V)
= \int_{V}  {\rd}^3 \bs{x}  
\( 
- v^2 \der_0 \phi  
  +\fr{1}{8\pi^2} 
\epsilon^{ijk}
a_i \der_j a_k 
\)\,.
\end{equation}
As $Q_{\phi}(V)$ is conserved and $|\der_0 \phi|$ decreases under the time evolution, the following quantity is generated as a consequence of the chiral instability:
\begin{equation}
\label{T(V)}
 T (V) = \int_V {\rd}^3 \bs{x}
\fr{1}{8\pi^2} \epsilon^{ijk} a_i \der_j a_k\,.
\end{equation}
This quantity is topological in that it does not depend on the spacetime metric. One can also show that it can be written in terms of a linking number~\cite{Moffatt1969}. 

To see the latter, consider temporally extended magnetic flux tubes with infinitesimal widths. Such a configuration can be represented by using delta functions,
\begin{equation}
\rd a = m_1 \delta_2 ({\cal S}_1; x) + m_2 \delta_2 ({\cal S}_2; x),
\label{da}
\end{equation}
where $m_{n}$ ($n=1,2$) are some constants and 
\begin{equation}
\delta_2 ({\cal S}_n;x)
:= \fr{\epsilon_{ \rho\sigma \mu\nu} }{2!2!}
\(\int_{{\cal S}_n} 
\rd y^\rho  \wed \rd y^\sigma 
 \delta^4 (x - y )\)
\rd x^\mu \wed \rd x^\nu\,,
\end{equation}
with ${\cal S}_{n}$ the worldsheets of the magnetic flux tubes located at the coordinates $(y^\mu)$.%
\footnote{One can easily see that \er{da} gives the configuration of the magnetic fluxes when ${\cal S}_n$ are temporally extended with the surface element $\rd y^0 \wed \rd y^i$ as
\begin{equation}
f_{jk} (x) = m 
\epsilon_{0i jk }
\int_{{\cal S}_n} \rd y^0 \wed \rd y^i
 \delta^4 (x - y )\,, 
\quad f_{0i} (x) =0.
\end{equation}
}
We assume that subspaces ${\cal S}_1$ and ${\cal S}_2$ do not have self-intersections. By using the property $\delta_2 ({\cal S}_n; x)  =  -\rd \delta_1 ({\cal V}_{{\cal S}_n};x)$ for a 3-dimensional subspace ${\cal V}_{{\cal S}_n}$ with the boundary ${\cal S}_n$,%
\footnote{This can be derived as follows (see, e.g., refs.~\cite{Chen:2015gma,Hidaka:2020izy}). For a 2-form field $\omega_2$, we have the following identity by the Stokes theorem:
\begin{equation}
\int \omega_2 \wed \delta_2 ({\cal S}_n; x)
=  
\int_{{\cal S}_n} \omega_2 
=
\int_{{\cal V}_{{\cal S}_n}} \rd \omega_2 
=
\int \rd \omega_2 \wed \delta_1 ({\cal V}_{{\cal S}_n};x)
=
- \int  \omega_2 \wed \rd \delta_1 ({\cal V}_{{\cal S}_n};x)\,,
\end{equation}
where the delta function $\delta_1 ({\cal V}_{{\cal S}_n}; x)$ defined by
\begin{equation}
 \delta_1 ({\cal V}_{{\cal S}_n}; x)
  = \fr{\epsilon_{\nu\rho\sigma \mu} }{3!}
\(\int_{{\cal V}_{{\cal S}_n}} \rd y^\nu \wed \rd y^\rho \wed \rd y^\sigma
\delta^4 (x-y)\) \rd x^\mu
\end{equation}
is only non-zero on ${\cal V}_{{\cal S}_n}$. This equation leads to the desired relation.}
we can express $a_\mu$ in terms of the delta function as
\begin{equation}
 a = - m_1 \delta_1 ({\cal V}_{{\cal S}_1}; x)
 - m_2 \delta_1 ({\cal V}_{{\cal S}_2}; x) 
\label{a_delta}
\end{equation}
up to a gauge transformation, which corresponds to the choice of the 3-dimensional subspaces ${\cal V}_{{\cal S}_1}$ and ${\cal V}_{{\cal S}_2}$.%
\footnote{For example, we can choose another 3-dimensional subspace ${\cal V}_{{\cal S}_1}'$ whose boundary is ${\cal S}_1$, and the gauge field can be expressed as $a' = - m_1 \delta_1 ({\cal V}_{{\cal S}_1}';x )
- m_2 \delta_1 ({\cal V}_{{\cal S}_2}; x)$.
The relation between $a$ and $a'$ is $a' = a - m_1 \delta_1 ({\cal V}_{{\cal S}_1}' \cup \bar{{\cal V}}_{{\cal S}_1};x )$.
Since ${\cal V}_{{\cal S}_1}' \cup \bar{{\cal V}}_{{\cal S}_1}$ is a closed subspace, it can be written as a boundary of 4-dimensional subspace 
$\Omega$ as 
$a' = a - m_1 \rd 
\delta_0 (\Omega)$,
where $\der \Omega = {\cal V}_{{\cal S}_1}'
\cup \bar{{\cal V}}_{{\cal S}_1}$.
Therefore, the change of the 3-dimensional subspace can be expressed by the gauge transformation with the gauge parameter $- m_I \delta_0 (\Omega)$.}

By substituting \er{da} and \er{a_delta} to \er{T(V)}, we have
\begin{align}
T (V)
&
 = - \fr{m_1m_2}{8\pi^2} 
\int_V 
\delta_1 ({\cal V}_{{\cal S}_1}) \wed \delta_2 ({\cal S}_2)
 - \fr{m_1m_2}{8\pi^2} 
\int_V
\delta_1 ({\cal V}_{{\cal S}_2}) \wed \delta_2 ({\cal S}_1)
\nonumber \\
&
= 
- \fr{m_1m_2}{4\pi^2} 
\int_V
\delta_1 ({\cal V}_{{\cal S}_1}) \wed \delta_2 ({\cal S}_2)\,,
\end{align}
where we performed the partial integral. Here and below, the delta function form $\delta_1 ({\cal V}_{{\cal S}};x )$ is abbreviated as $\delta_1 ({\cal V}_{{\cal S}})$. The integral in the second line is only non-zero where all of $V$, ${\cal V}_{{\cal S}_1}$, and ${\cal S}_2$ intersect transversally, and gives $\pm 1$ on the intersection points according to their orientations. Since $V$ is assumed to be a time slice, the integral reduces to the intersection number between ${\cal V}_{{\cal S}_1}$ and ${\cal S}_2$ on $V$, which is equal to the linking number between ${\cal S}_1$ and ${\cal S}_2$ on $V$:
\begin{equation}
\link ({\cal S}_1, {\cal S}_2; V)
:= 
\int_V
\delta_1 ({\cal V}_{{\cal S}_1}) \wed \delta_2 ({\cal S}_2)\,.
\end{equation}
We thus arrive at 
\begin{equation}
  T (V)
 = - \fr{m_1m_2}{4\pi^2} \link ({\cal S}_1, {\cal S}_2; V)\,.
\end{equation}

Before ending this section, we provide a physical argument for this instability and the creation of the configuration of magnetic flux with non-trivial linking~\cite{Akamatsu:2014yza}.

\begin{figure}[t]
\begin{center}
 \ig[width=37.5em]{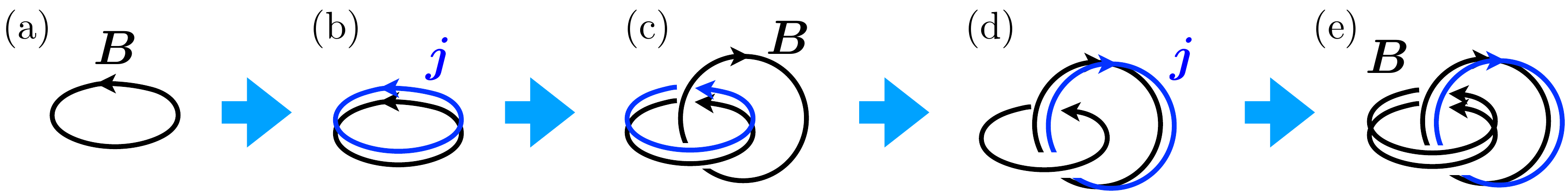} 
\end{center}
\caption{\label{chinst} Schematic picture of the chiral instability~\cite{Akamatsu:2014yza}. The magnetic fields and electric currents are expressed as black and blue closed loops, respectively. The induced electric currents parallel to the magnetic fields lead to the growth of the magnetic fields.}
\end{figure}

We begin with the system with the initial configuration $\der_0 \b \phi \neq 0$ and consider a fluctuation of a small magnetic flux tube (figure~\ref{chinst}(a)). This small magnetic flux induces an electric current due to the chiral magnetic effect in \er{CME} (figure~\ref{chinst}(b)). By the Amp\`ere law, the induced current creates a magnetic flux linked with the original magnetic flux (figure~\ref{chinst}(c)). By the chiral magnetic effect again, we have an additional electric current (figure~\ref{chinst}(d)). This current amplifies the original small fluctuation by the Amp\`ere law (figure~\ref{chinst}(e)).%
\footnote{Precisely speaking, a static magnetic field configuration (without the amplification) is possible when the condition $\bs{\na}\times \bs{B} = \b \sigma \bs{B}$ is satisfied. This field corresponds to the Beltrami field with the wave number $|\bs{k}|=|\bar \sigma|$ as described in section~\ref{CPImode}. When $|\bs{k}|<|\bar \sigma|$, the static configuration cannot be sustained, and the amplification of the magnetic field occurs.} 
Therefore, the small fluctuation of the magnetic flux becomes unstable, and we end up with the configuration of the magnetic flux with non-trivial linking.

\begin{figure}[t]
    \centering
\includegraphics[height=5em]{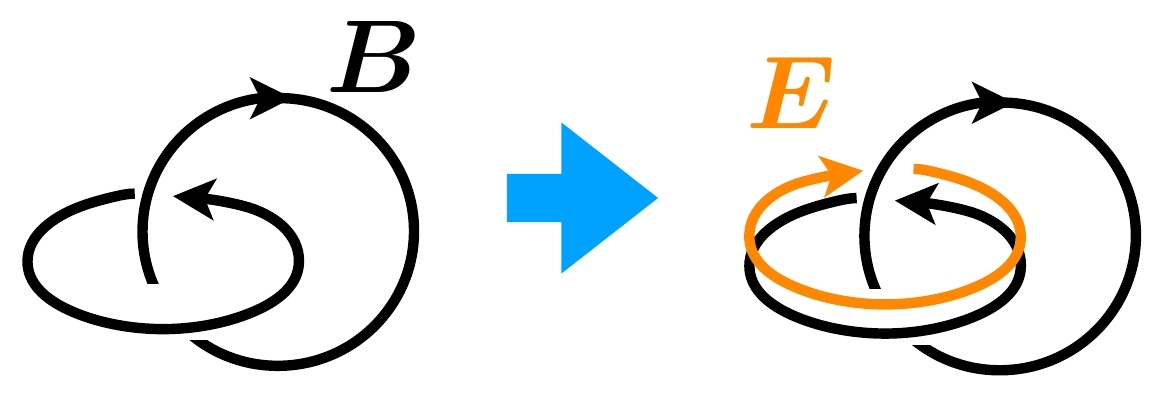}
\caption{\label{Faraday}Induced electric flux due to the Faraday law.
The electric field is described by the orange closed loop. The direction of the electric flux is opposite to that of the magnetic flux, so $\bs{E}\cdot \bs{B} < 0$.}
\end{figure}

On the other hand, the growth of the magnetic flux causes an induced electric flux by the Faraday law, $\bs{\na}\times \bs{E} = - \der_0 \bs{B}$. It suggests that the electric flux direction is opposite to the magnetic flux (figure~\ref{Faraday}). 
As a consequence, $|\der_0 \phi|$ is reduced by the equation of motion for $\phi$, i.e., $v^2 \der_0^2 \phi = \fr{1}{4\pi^2} \bs{E} \cdot \bs{B} < 0$.

\subsection{\label{CPInoninv}Magnetic helicity and non-invertible symmetries}

In the above discussion, we considered the conserved charges in \ers{Q_phi} and \eqref{Q_a}, which contain the terms $\int_{\cal V} a \wed \rd a$ and $\int_{\cal S} \phi \rd a$, respectively. While these terms are invariant under small gauge transformations under certain boundary conditions, they are not invariant under large gauge transformations. In particular, the former is called the magnetic helicity and its role has been intensively studied in the context of plasma physics and magnetohydrodynamics~\cite{Davidson2001}, but the lack of large gauge invariance has not been fully appreciated. Here, we provide the gauge-invariant formulation of these conserved charges following refs.~\cite{Choi:2022jqy,Cordova:2022ieu,Choi:2022fgx,Yokokura:2022alv}, where the gauge-invariant conserved charges are identified as symmetry defects for non-invertible symmetries. 

First, we see that 
\begin{align}
 \re^{\ri \alpha_\phi  Q_\phi ({\cal V})} 
&=
\re^{\ri \alpha_\phi \int_{\cal V} (v^2 * \rd \phi + \fr{1}{8\pi^2} a \wed \rd a)},
\\
\re^{\ri \alpha_a  Q_a ({\cal S})}
&= 
\re^{\ri \alpha_a 
\int_{\cal S} (-\fr{1}{e^2} * \rd a  + \fr{1}{4\pi^2} \phi \rd a)},
\end{align}
constructed from the conserved charges violate the large gauge invariance due to the topological terms $\int_{\cal V} a \wed \rd a$ and $\int_{\cal S} \phi \rd a$. 

To make the integrands gauge invariant, we define the topological terms  $\re^{\ri\alpha_\phi \int_{\cal V}  \fr{1}{8\pi^2} a \wed \rd a}$ and $\re^{\ri \alpha_a \int_{\cal S} \fr{1}{4\pi^2} \phi \rd a}$ by using auxiliary 4- and 3-dimensional spaces $\Omega_{\cal V}$ and ${\cal V_S}$ whose boundaries are ${\cal V}$ and ${\cal S}$ as
\begin{align}
\re^{\ri \alpha \int_{\cal V}  \fr{1}{8\pi^2} a \wed \rd a}
&:=
\re^{\ri \alpha \int_{\Omega_{\cal V}}  \fr{1}{8\pi^2} \rd a \wed \rd a},
\\
 \re^{\ri \alpha_a \int_{\cal S} \fr{1}{4\pi^2} \phi \rd a}
&:= 
\re^{\ri \alpha_a \int_{\cal V_S} \fr{1}{4\pi^2}  \rd\phi \wed \rd a},
\end{align}
respectively.
The integrands are now manifestly gauge invariant, but we have chosen the spaces $\Omega_{\cal V}$ and ${\cal V_S}$ by hand. The absence of the ambiguity in the choice of $\Omega_{\cal V}$ and ${\cal V_S}$ requires 
\begin{gather}
\re^{\ri \alpha \int_{\Omega_{\cal V}} \fr{1}{8\pi^2} \rd a \wed \rd a}
\re^{- \ri \alpha \int_{\Omega_{\cal V}'} \fr{1}{8\pi^2} \rd a \wed \rd a}
=
\re^{\ri \alpha \int_{\Omega_{\cal V} \cup \b\Omega_{\cal V}'} 
\fr{1}{8\pi^2} \rd a \wed \rd a}
= 1,
\\
\re^{\ri \alpha_a \int_{\cal V_S} \fr{1}{4\pi^2}  \rd\phi \wed \rd a}
\re^{- \ri \alpha_a \int_{\cal V_S'}
\fr{1}{4\pi^2}  \rd\phi \wed \rd a} = 
\re^{ \ri \alpha_a \int_{{\cal V_S} \cup \bar{\cal V}'_{\cal S} }
 \fr{1}{4\pi^2}  \rd\phi \wed \rd a}
=1.
\end{gather}
Here, $\Omega_{\cal V}'$ and ${\cal V}_{\cal S}'$ are other 4- and 3-dimensional spaces with the boundaries ${\cal V}$ and ${\cal S}$, and $\bar\Omega_{\cal V}'$, and $\bar{\cal V}_{\cal S}'$ denote $\Omega_{\cal V}'$ and ${\cal V}_{\cal S}'$ with opposite orientations, respectively.
By using the flux quantization conditions $\int_{\Omega_{\cal V} \cup \b\Omega_{\cal V}'}\rd a \wed \rd a \in 8\pi^2 \bb{Z}$ and $\int_{{\cal V_S }\cup \bar{\cal V}_{\cal S}'} \rd\phi \wed \rd a \in 4\pi^2 \bb{Z}$, we require $\re^{\ri\alpha_\phi} =1$ and $\re^{\ri\alpha_a} =1$, where we have used the fact that $\Omega_{\cal V} \cup \b\Omega_{\cal V}'$ and ${\cal V_S }\cup \bar{\cal V}_{\cal S}'$ are 4- and 3-dimensional spaces without boundaries, respectively. These requirements mean that $\re^{\ri \alpha_\phi Q_\phi ({\cal V})}$ and $\re^{\ri \alpha_a Q_a ({\cal S})}$ are just trivial topological operators that do not generate symmetry transformations.

There is alternative way to construct the gauge-invariant topological operators that generate non-trivial symmetry transformations. If the parameters $\alpha_\phi $ and $\alpha_a$ are $2\pi$ multiples of rational numbers, we can regard the topological terms as naive partition functions of topological quantum field theories. For simplicity, we consider the case where the parameters $\alpha_\phi$ and $\alpha_a$ are given by $\alpha_\phi = \fr{2\pi}{q_\phi}$ and $\alpha_a  = \fr{2\pi}{q_a}$ with $q_\phi, q_a \in \bb{Z}$. In this case, we can modify the topological terms $\re^{\fr{\ri}{4\pi q_\phi} \int_{\cal V} a \wed \rd a}$ and $\re^{\fr{\ri}{2\pi q_a} \int_{\cal S} \phi \rd a}$ by introducing a $2\pi$ periodic scalar field $\chi_a$, 1-form gauge fields $c_\phi$ and $c_a$ with the flux quantization conditions $\int_{\cal C} \rd \chi_a, \int_{\cal S} \rd c_\phi, \int_{\cal S} \rd c_a \in 2\pi \bb{Z}$ as
\begin{gather}
\re^{\fr{\ri}{4\pi q_\phi} \int_{\cal V}a \wed \rd a}
\to 
\int {\cal D} c_\phi \, 
\re^{\fr{\ri}{4\pi} \int_{\cal V}
(-q_\phi c_\phi \wed \rd c_\phi + 2 c_\phi \wed \rd a)
},
\\
 \re^{\fr{\ri}{2\pi q_a} \int_{\cal S} \phi \rd a}
 \to 
\int {\cal D} [\chi_a ,c_a] \, 
\re^{\fr{\ri}{2\pi} \int_{\cal S}
(-q_a \chi_a \rd c_a + \chi_a \rd a + \phi \rd c_a)
}.
\label{non-invert_phase}
\end{gather}
The right-hand sides do not depend on the choice of auxiliary spaces because $q_\phi$ and $q_a$ are now in the numerators.%
\footnote{One may naively integrate out $c_\phi$ by substituting the equation of motion $q_\phi \rd c_\phi = \rd a$ to obtain the naive action $\re^{\fr{\ri}{4\pi q_\phi } \int_{\cal V}a \wed \rd a}$.
However, this naive solution would violate the flux quantization condition except for the case $\int_{\cal S} \rd a \in 2\pi q_a \bb{Z}$. A similar remark applies to the naive action $\re^{\fr{\ri}{2\pi q_a} \int_{\cal S} \phi \rd a}$ obtained by integrating out $\chi_a$ and $c_a$ in \er{non-invert_phase}.}
Therefore, we can modify $\re^{\fr{2\pi \ri}{q_\phi } Q_\phi ({\cal V})}$ and $\re^{\fr{2\pi \ri}{q_a } Q_a({\cal S})}$ as
\begin{equation}
\re^{\fr{2\pi \ri}{q_\phi } Q_\phi ({\cal V})}
\to 
D_\phi (\re^{2\pi \ri /q_\phi },{\cal V})
:= 
\re^{\fr{2\pi \ri}{q_\phi } \int_{\cal V} v^2 * \rd \phi}
\int {\cal D} c_\phi \, 
\re^{\fr{\ri }{4\pi} \int_{\cal V}
(- q_\phi c_\phi  \wed \rd c_\phi + 2 c_\phi \wed \rd a)
},
\end{equation}
and 
\begin{equation}
\re^{\fr{2\pi \ri}{q_a } Q_a ({\cal S})}
\to 
D_a (\re^{2\pi \ri /q_a },{\cal S})
:= 
\re^{ - \fr{2\pi \ri}{q_a } \int_{\cal S} \fr{1}{e^2} * \rd a}
\int {\cal D} [\chi_a, c_a] \, 
\re^{\fr{\ri }{2\pi} \int_{\cal S}
(- q_a \chi_a \rd c_a + \chi_a \rd a + \phi \rd c_a)
}, 
\end{equation}
respectively. These objects are gauge-invariant, but non-invertible because their expectation values are not equal to unity depending on the topology of ${\cal V}$ and ${\cal S}$.

\begin{figure}[t]
\begin{center}
 \includegraphics[width=35.5em]{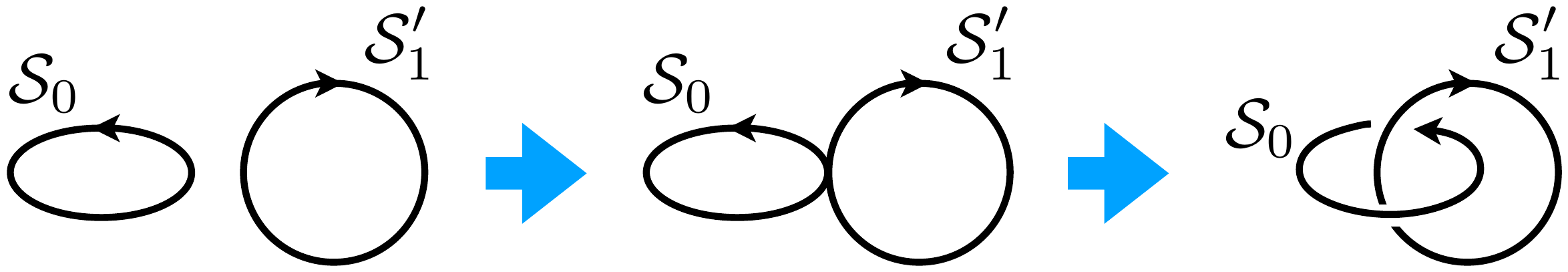} 
\end{center}
\caption{\label{noninvlink}
A time series of the worldsheets ${\cal S}_0$ and ${\cal S}_1'$. 
The magnetic flux tubes on ${\cal S}_0$ and ${\cal S}_1'$ are created without linking (left panel), intersected at some time (middle panel), and  linked after the intersection (right panel).}
\end{figure}

We can now show the relation between the magnetic helicity and the linking number from the viewpoint of the non-invertible symmetries. Our strategy is to construct a linked configuration of the magnetic flux tubes from an unlinked one. Noting that the defect $D_a (\re^{\fr{2\pi \ri}{q_a}},{\cal S})$ may be thought of as a magnetic flux tube on the worldsheet ${\cal S}$, we begin with the correlation function of two isolated (unlinked) flux tubes, $\vevs{D_a (\re^{\fr{2\pi \ri}{q_a}}, {\cal S}_0) D_a (\re^{\fr{2\pi \ri}{q'_a}}, {\cal S}'_0)}$, whose expectation value is non-zero in general. By taking ${\cal S}_0$ and ${\cal S}'_0$ as temporally extended surfaces, the symmetry defects represent the worldsheets of magnetic flux tubes. Then, we continuously move ${\cal S}'_0$ to ${\cal S}'_1$, so that ${\cal S}'_1$ is linked with ${\cal S}_0$ on a time slice (see figure~\ref{noninvlink}).

The continuous deformation can be done by using a 3-dimensional subspace 
${\cal V}'_{01} $ whose boundary is given by $\der {\cal V}'_{01} = {\cal S}_0' \sqcup \bar{\cal S}_1'$. By the continuous deformation, we have the following relation:
\begin{align}
&
 \vevs{D_a (\re^{\fr{2\pi \ri}{q_a}}, {\cal S}_0)
D_a (\re^{\fr{2\pi \ri}{q'_a}}, {\cal S}'_0)}
\nonumber \\
&
=  
\int {\cal D}[\phi, a]
D_a (\re^{\fr{2\pi \ri}{q_a}}, {\cal S}_0)
\re^{\ri S [\phi, a]} 
\re^{-\fr{2\pi \ri}{q_a'}\int_{{\cal S}_0'} \fr{1}{e^2}* \rd a}
\int {\cal D}[\chi_a', c_a']
\re^{\fr{\ri}{2\pi} \int_{{\cal S}_0'}  (- q'_a \chi_a' \rd c_a'  
+ \chi'_a \rd a  + \phi \rd c_a')}
\nonumber \\
&
=  
\int {\cal D}[\phi, a]
D_a (\re^{\fr{2\pi \ri}{q_a}}, {\cal S}_0)
D_a (\re^{\fr{2\pi \ri}{q'_a}}, {\cal S}'_1)
\re^{\ri S [\phi, a]} 
\re^{-\fr{2\pi \ri}{q_a'} 
\int \fr{1}{e^2}* \rd a \wed ( - \rd \delta_1 ({\cal V}_{01}'))}
\nonumber \\
&
\quad 
\times 
\int {\cal D}[\chi_a', c_a']
\re^{ \fr{\ri}{2\pi} \int  (- q'_a \chi_a' \rd c_a'  
+ \chi_a' \rd a  + \phi \rd c'_a) 
\wed ( - \rd \delta_1 ({\cal V}_{01}'))},
\end{align}
where we used $\delta_2 ({\cal S}'_0) =  \delta_2 ({\cal S}'_1) - \rd \delta_1 ({\cal V}_{01}')$. Since there are no singularities (such as axion vortices and magnetic monopoles) on ${\cal V}_{01}'$, we can safely integrate out $\chi_a$ and $c_a$ on ${\cal V}_{01}'$ to obtain
\begin{align}
&
\vevs{D_a (\re^{\fr{2\pi \ri}{q_a}}, {\cal S}_0)
D_a (\re^{\fr{2\pi \ri}{q'_a}}, {\cal S}'_0)}
\nonumber \\
&
=  
\int {\cal D}[\phi, a]
D_a (\re^{\fr{2\pi \ri}{q_a}}, {\cal S}_0)
D_a (\re^{\fr{2\pi \ri}{q'_a}}, {\cal S}'_1)
\re^{\ri S [\phi, a]} 
\re^{-\fr{2 \pi \ri}{q'_a}
\int \fr{1}{e^2}* \rd a \wed ( - \rd \delta_1 ({\cal V}_{01}'))}
\re^{\fr{\ri}{2\pi q'_a} \int_{{\cal V}_{01}'} \rd\phi \wed \rd a}.
\end{align}
We can absorb the last two exponential factors on ${\cal V}_{01}'$ into the factor $\re^{\ri S [\phi, a]}$ by the change of the variable $a \to a - \fr{2\pi}{q_a'} \delta_1 ({\cal V}_{01}')$. Then, $D_a (\re^{\fr{2\pi \ri}{q_a}}, {\cal S}_0)$, which can be defined on a 3-dimensional subspace ${\cal V}_{{\cal S}_0}$ with the boundary ${\cal S}_0$, receives the following transformation:
\begin{equation}
D_a (\re^{\fr{2\pi \ri}{q_a}}, {\cal S}_0)
\to 
\re^{-\fr{1}{e^2}\int_{{\cal S}_0}  * \rd a}
\int {\cal D}[\chi_a, c_a]
\re^{-\fr{\ri}{q_a'} 
\int_{{\cal V}_{{\cal S}_0}} \rd \chi_a \wed \rd \delta_1 ({\cal V}_{01}')}
\re^{\fr{\ri}{2\pi} \int_{{\cal V}_{{\cal S}_0}} (- q_a \rd \chi_a \wed 
\rd c_a + \rd \chi_a \wed \rd a + \rd \phi \wed \rd c_a)}.
\end{equation}
By further redefining $c_a + \fr{2\pi}{q_a  q_a'} \delta_1 ({\cal V}_{01}')\to c_a$, we have 
\begin{equation}
D_a (\re^{\fr{2\pi \ri}{q_a}}, {\cal S}_0)
\to
D_a (\re^{\fr{2\pi \ri}{q_a}}, {\cal S}_0)
\re^{- \fr{\ri}{q_a q_a'} \int_{{\cal V}_{{\cal S}_0}} 
\rd \phi \wed \rd \delta_1 ({\cal V}_{01}')}  
=
D_a (\re^{\fr{2\pi \ri}{q_a}}, {\cal S}_0)
\re^{- \fr{\ri}{q_a q_a'} \int_{{\cal V}_{{\cal S}_0}} 
\rd \phi \wed \delta_2 ({\cal S}'_1)}.
\end{equation}
Therefore, we obtain 
\begin{equation}
  \vevs{D_a (\re^{\fr{2\pi \ri}{q_a}}, {\cal S}_0)
D_a (\re^{\fr{2\pi \ri}{q'_a}}, {\cal S}'_0)}
=  \vevs{\re^{- \fr{\ri}{q_a q_a'} \int_{{\cal V}_{{\cal S}_0}} 
\rd \phi \wed \delta_2 ({\cal S}'_1)}
D_a (\re^{\fr{2\pi \ri}{q_a}}, {\cal S}_0)
D_a (\re^{\fr{2\pi \ri}{q'_a}}, {\cal S}'_1)}.
\end{equation}

To relate the magnetic helicity to the linking number, we consider the 0-form symmetry defect $D_\phi (\re^{\fr{2\pi \ri }{q_\phi}},{\cal V})$ that contains a gauge-invariant expression of the magnetic helicity. We insert $D_\phi (\re^{\fr{2\pi \ri}{q_\phi }},{\cal V}_0)$ into the above correlation, $\vevs{D_\phi (\re^{\fr{2\pi \ri}{q_\phi} },{\cal V}_0) D_a (\re^{\fr{2\pi \ri}{q_a}}, {\cal S}_0) D_a (\re^{\fr{2\pi \ri}{q'_a}}, {\cal S}'_0)}$, where ${\cal V}_0$ does not intersect ${\cal S}_0$ and ${\cal S}_0'$. After the continuous deformation from ${\cal S}_0 '$ to ${\cal S}_1'$, we continuously deform from ${\cal V}_0 $ to ${\cal V}_1$ that intersects ${\cal S}_0$ and ${\cal S}'_1$. This can be done by using 4-dimensional subspace $\Omega_{01}$ satisfying $\der\Omega_{01} = {\cal V}_0 \sqcup {\cal V}_1$. By the similar procedure and the shift $\phi \to \phi + \fr{2\pi}{q_\phi } \delta_0 (\Omega_{01})$, we obtain the following relation:
\begin{align}
&
  \vevs{D_\phi (\re^{\fr{2\pi \ri}{q}},{\cal V}_0) 
D_a (\re^{\fr{2\pi \ri}{q_a}}, {\cal S}_0)
D_a (\re^{\fr{2\pi \ri}{q'_a}}, {\cal S}'_0)}
\nonumber \\
&
= 
\re^{- \fr{2\pi \ri}{q_\phi  q_a q_a'} \int 
\delta_1 ({\cal V}_1) 
\wed \delta_2 ({\cal S}'_1) \wed \delta_1 ({\cal V}_{{\cal S}_0})}
\nonumber \\
&
\quad 
\times 
\langle
\re^{- \fr{\ri}{q_a q_a'} \int_{{\cal V}_{{\cal S}_0}} 
\rd \phi \wed \delta_2 ({\cal S}'_1)}
\re^{- \fr{\ri}{q_\phi q_a} \int 
\rd a 
\wed 
\delta_1 ({\cal V}_1)
\wed 
\delta_1 ({\cal V}_{{\cal S}_0}) 
}
\re^{- \fr{\ri}{q_\phi q_a'} \int 
\rd a 
\wed 
\delta_1 ({\cal V}_1)
\wed 
\delta_1 ({\cal V}_{{\cal S}'_1}) 
}
\nonumber \\
&
\qquad 
\times 
D_\phi (\re^{\fr{2\pi \ri}{q_\phi }},{\cal V}_1) 
D_a (\re^{\fr{2\pi \ri}{q_a}}, {\cal S}_0)
D_a (\re^{\fr{2\pi \ri}{q'_a}}, {\cal S}'_1)\rangle 
\,.
\end{align}
The factor $\re^{-\fr{2\pi \ri}{q_\phi q_a q_a'} \int \delta_1 ({\cal V}_1) \wed \delta_2 ({\cal S}'_1) \wed \delta_1 ({\cal V}_{{\cal S}_0})}$ implies the values of the magnetic fluxes $\fr{2\pi}{q_a}$ and $\fr{2\pi}{q_a'}$ with the linking number of ${\cal S}'_1$ and ${\cal S}_0$ on ${\cal V}_1$.

\section{\label{GCI}Generalized chiral instabilities}

In this section, we generalize the chiral instability in the electrodynamics above to instabilities for general $p$-form gauge theories  in $D$-dimensional Minkowski spacetime, which we call generalized chiral instabilities. Here, the time derivative of the axion field (or the chiral chemical potential) is generalized to electric fields of $p$-form gauge fields. In this section, we call the $(p+1)$-form electric and magnetic fields simply the electric and magnetic fields unless otherwise stated.

\subsection{Unstable modes}

We consider a higher-form gauge theory described by the following action:
\begin{equation}
 S = 
-
\fr{F^2_{IJ}}{2}
\int 
 {\rd}a_I \wed *{\rd} a_J
+ 
\fr{1}{3!} 
\int M_{IJK} a_I \wed {\rd}a_J \wed  {\rd} a_K,
\end{equation}
where $a_I$ ($I= 1,...,N$) are $p_I$-form gauge fields with the flux quantization conditions $\int_{\Sigma_{p_I +1}} \rd a_I \in 2\pi \bb{Z}$ for a closed $(p_I+1)$-dimensional subspace $\Sigma_{p_I +1}$, $F_{IJ}^2 = F_{IL} F_{JL}$ is a positive definite symmetric matrix with $F_{IJ}$ being an invertible constant matrix (which may be regarded as a generalization of the inverse of a coupling constant matrix), and $M_{IJK}$ is a constant tensor satisfying 
\begin{equation}
M_{IJK} = (-1)^{(p_I+1)(p_J+1)} M_{JIK}=(-1)^{(p_J +1)(p_K +1) } M_{I K J},
\end{equation}
so that the topological term does not vanish identically. Further, the components of $M_{IJK}$ are only non-zero if $p_I + p_J + p_K +2 = D$. The flux quantization condition requires that the components of $ M_{IJK} $ should be $\fr{1}{4\pi^2}$ multiples of integers.
We assume that the spacetime is a spin manifold.

The equation of motion for $a_I$ is given by
\begin{equation}
 (-1)^{p_I+1} F^2_{IJ} \rd * \rd a_J 
 =\fr{1}{2!}
M_{IJK} \rd a_J \wed \rd a_K\,.
\label{EOM_aI}
\end{equation}
First, we show that there are unstable modes in the case where some of gauge fields have the initial configuration,
\begin{equation}
\rd  \bar{a}_K = \fr{1}{p_K!}
\bar{f}_{K, 0 i_1....i_{p_K}} \rd x^0 \wed \rd x^{i_1} \weds \rd x^{i_{p_K}}\,.
\end{equation}
Here $\bar{f}_{K, 0 i_1....i_{p_K}} = - \bar E_{K, i_1....i_{p_K}}$ is a constant, $\der_\mu  \bar E_{K, i_1....i_{p_K}} =0$. 
We decompose the gauge field as $a_I = \bar{a}_I +\delta a_I$, where $\bar{a}$ is the initial configuration and $\delta a$ is the fluctuation. The equation of motion can be obtained by replacing $\rd a_I \to \rd \bar{a}_I+ \rd \delta a_I$ as
\begin{equation}
(-1)^{p_I +1}  F^2_{IJ} \rd * \rd \delta a_J =
M_{IJK} \left( \rd \delta a_J \wed \rd \bar{a}_K
+
\fr{1}{2!} \rd \delta a_J \wed \rd \delta a_K \right)\,,
\end{equation}
where we have used $\rd \bar{a}_J \wed \rd \bar{a}_K \propto \rd x^0 \wed \rd x^0  = 0$.

Let us now see the existence of unstable modes by the linear analysis. For this purpose, we perform the Fourier decomposition for the fluctuation $\delta a$ as $\hat{a}_{\cal I} (t, \bs{k})
= F_{IJ}
 \int  \rd^{D-1} \bs{x} 
\delta a_{J,i_1...i_{p_I} } (t, \bs{x}) 
\re^{- \ri\bs{k} \cdot \bs{x}}$,
where we introduced an unified notation ${\cal I} = (I, i_1,...,i_{p_I})$ (e.g., $a_{\cal I} = a_{I, i_1...i_{p_I}}$) with the ordering $i_1 < i_2< \cdots < i_{p_I}$ to avoid overcounting.
We included the factor $F_{IJ}$ on the right-hand side so that the equation of motion is canonically normalized. The Fourier mode $\hat{a}_{\cal I} (t,\bs{k})$ satisfies the reality condition $\hat{a}_{\cal I}^* (t,\bs{k}) = \hat{a}_{\cal I} (t,-\bs{k})$.
The equation of motion in momentum space can then be rewritten as
\begin{equation}
 (\omega^2 - |\bs{k}|^2) \hat{a}_{\cal I} 
= \ri k_l \hat{M}_{\cal IJ}^l \hat{a}_{\cal J},
\label{EOM_aI_momentum}
\end{equation}
where we neglected the non-linear terms and took the temporal gauge 
$a_I^{0 i_1....i_{p_I-1}} =0$ with the constraint $k_i a_I^{i i_1...i_{p_I-1}} =0$. The matrix $\hat{M}_{\cal IJ}^l$ represents the homogeneous electric field normalized by $F_{IJ}$,
\begin{equation}
 \hat{M}_{\cal IJ}^l
= \fr{\epsilon^{i_1...i_{p_I} l j_1...j_{p_J} 
0 k_1...k_{p_{K}}}}{p_{K}!}
F^{-1}_{IK} F^{-1}_{JL} M_{KL M } \bar{f}_{M, 0 k_1...k_{p_{K}}}\,.
\end{equation}

Since $ \hat{M}_{\cal IJ}^l$ is anti-symmetric under the exchange ${\cal I} \corr {\cal J}$, we can block diagonalize $\hat{M}_{\cal IJ}^l$ using a real orthogonal matrix $P_{\cal IJ}$ as 
\begin{equation}
 P_{\cal IK}^{\rm T} \hat{M}_{\cal KL}^l k_l P_{\cal LJ} = 
\Lambda_{\cal IJ},
\label{P}
\end{equation}
where
\begin{equation}
\Lambda_{\cal IJ}
:= 
 \mtx{
\ri \sigma_2 \lambda_1 
\\
& \ddots
\\
&& \ri \sigma_2 \lambda_{\cal N}
\\
&&& 0
}_{\cal IJ}\,,
\end{equation}
with $\sigma_2$ being the second Pauli matrix, $\lambda_n (\bs{k})$ ($n = 1, ..., {\cal N}$) the positive eigenvalues of $M_{\cal IJ}^l k_l$, and ${\cal N} = \fr{1}{2} \rank (M_{\cal IJ}^l k_l)$.
The dispersion relation for each block diagonal sector can be derived as
\begin{equation}
 \omega^2 = |\bs{k}|^2 \pm \lambda_n ,
\end{equation}
where the modes with the negative sign, $\omega^2 = |\bs{k}|^2 - \lambda_n$, become tachyonic in the infrared region, $0 <  |\bs{k}|^2<  \lambda_n$. Therefore, there are ${\cal N}$ unstable modes in the presence of the initial homogeneous electric field~\cite{Yamamoto:2022vrh}.

From this argument, we can introduce a higher-form generalization of the Beltrami field. It is given as the $p$-form gauge field at the critical value $|\bs{k}|^2 = \lambda_n$ satisfying the relation $F^2_{IJ} \na^2 \delta a_{J , i_1...i_p} = M^{l}_{\cal IK} \der_l \delta a_{\cal K} $. Here, the matrix $ M^{l}_{\cal IK} $ is a generalization of $-\b\sigma \epsilon_{ilk}$ for the Beltrami field in the case of the chiral instability in section~\ref{CPImode}.

\subsection{Reduction of electric fields}

Let us show that the unstable modes tend to reduce the initial electric fields. We assume the initial electric fields $\bar{f}_I^{0 i_1...i_{p_I}} = \bar{E}_I^{i_1...i_{p_I}}$ for some of the gauge fields.
To compute the time evolution of the electric fields, we consider the equation of motion in \er{EOM_aI}. In particular, the higher-form generalization of the Maxwell-Amp\`ere law is given by 
\begin{align}
& 
F^2_{IJ} \der_0 f_J^{0 i_1...i_{p_J}} 
+
F^2_{IJ} \der_k  f_J^{k i_1... i_{p_J}}
\nonumber \\
&=
\fr{1 
}{2!(p_J +1)! (p_K+1)!}
M_{IJK} \epsilon^{ i_1 ...i_{p_I} \nu \nu_1...\nu_{p_J} 
\rho \rho_1...\rho_{p_K}}
 f_{J, \nu \nu_1...\nu_{p_J}} f_{K, \rho \rho_1...\rho_{p_K}}\,. 
\end{align}
By multiplying both hand sides by $\bar{f}_{I,0 i_1...i_{p_I}} = - \bar{E}_{i_1...i_{p_I}}$ and integrating over the space, we have 
\begin{equation}
\int \rd^{D-1} \bs{x}
F^2_{IJ}
\bar{E}_{I, i_1...i_{p_I}} 
\der_0 
\delta f_J^{0 i_1...i_{p_I}}
=
\int \rd^{D-1} \bs{x}
M_{\cal IJ}^l 
\der_0 
\delta a_{\cal I}
\der_l \delta a_{\cal J} 
\,. 
\label{EOM_aI2}
\end{equation}
Here, we have decomposed the field strength $f_{I,0 i_1...i_{p_I}}$ into the initial configuration and the fluctuation, $f_{I,0 i_1...i_{p_I}} = \bar{f}_{I,0 i_1...i_{p_I}} + \delta f_{I,0 i_1...i_{p_I}}$, and dropped surface terms.
Below we will show that the right-hand side of eq.~(\ref{EOM_aI2}) is negative, and since the integrand on the left-hand side includes the time derivative of the summation of the electric fields, at least one electric field must be antiparallel to the initial electric field such that the total electric field is reduced.

To evaluate the right-hand side, it is convenient to go to momentum space as
\begin{equation}
 \int {\rd}^{D-1} \bs{x}
 M_{\cal IJ}^l 
 \der_0 \delta a_{\cal I}
 \der_l \delta a_{\cal J}
= 
\int \fr{{\rd}^{D-1} \bs{k}}{(2\pi)^{D-1}}
\der_0 \hat{a}_{\cal I}(t, -\bs{k})
\, 
\ri \hat{M}_{\cal IJ}^l k_l \hat{a}_{\cal J} (t, \bs{k})\,,
\end{equation}
where we used that $M_{\cal IJ}^l$ does not depend on the spatial coordinates. By using the matrices $\Lambda_{\cal IJ}$ and $P_{\cal IJ}$ given in \er{P}, we have
\begin{equation}
\der_0 \hat{a}_{\cal I}(t, -\bs{k})
\, 
\ri 
 \hat{M}_{\cal IJ}^l k_l \hat{a}_{\cal J} (t, \bs{k})
= 
\der_0
 v_{\cal I}(t, -\bs{k})
\, 
\ri 
\Lambda_{\cal IJ}
 v_{\cal J}(t, \bs{k}) ,
\label{E_evolution}
\end{equation}
where $v_{\cal J}$ is a vector satisfying $\hat{a}_{\cal I} = P_{\cal IJ} v_{\cal J}$, and the matrix $P_{\cal IJ}$ is introduced in \er{P}.
We note that the positive eigenvalues $\lambda_1 (\bs{k}),..., \lambda_n (\bs{k})$ are invariant under $\bs{k} \to -\bs{k}$, since this operation is equivalent to $M_{\cal IJ}^l k_l \to M_{\cal JI}^l k_l$.

To evaluate \er{E_evolution}, we use the solution of $v_{\cal J}$. We can diagonalize the equation of motion in \er{EOM_aI_momentum} as
\begin{equation}
(-\der_0^2 - |\bs{k}|^2) v_{\cal I} (t,\bs{k}) 
=  \ri \Lambda_{\cal IJ}v_{\cal J} (t,\bs{k}) .
\end{equation}
This equation can be solved as
\begin{align}
 v_{\cal J} (t,\bs{k}) 
&
 = 
\sum_{n} 
\epsilon^{\rm L}_{{\cal J},n}
\(A^{+}_{\bs{k},\lambda_n} {\re}^{{\ri} t  \sr{|\bs{k}|^2 +\lambda_n} }
 + A^{-}_{\bs{k},\lambda_n} 
{\re}^{-{\ri} t \sr{|\bs{k}|^2 +\lambda_n}}
\)
\nonumber \\
&\quad
+
\sum_{n, \lambda_n < |\bs{k}|^2}
\epsilon^{\rm R}_{{\cal J},n}
\(
A^{+}_{\bs{k},-\lambda_n} {\re}^{{\ri} t  \sr{|\bs{k}|^2 -\lambda_n} }
 + A^{-}_{\bs{k},-\lambda_n}
 {\re}^{-{\ri} t \sr{|\bs{k}|^2 -\lambda_n}}
\)
\nonumber \\
&\quad
+
\sum_{n, \lambda_n > |\bs{k}|^2}
\epsilon^{\rm R}_{{\cal J},n}
\(A^{+}_{\bs{k},-\lambda_n} {\re}^{ t \sr{\lambda_n - |\bs{k}|^2} }
 + A^{-}_{\bs{k},-\lambda_n} 
 {\re}^{- t \sr{\lambda_n  - |\bs{k}|^2}}\)
+\cdots\,,
\label{v_J}
\end{align}
where 
\begin{equation}
 \epsilon^{h}_{{\cal J},n} = 
\fr{1}{\sr{2}}
(\underbrace{0,...,0}_{2n-2},{1, \mp \ri},\underbrace{0,...,0}_{2({\cal N} - n )})
\end{equation}
are polarization vectors for $h = {\rm L,R}$. In \er{v_J}, we have ignored the zero-eigenvalue modes that do not contribute to \er{E_evolution}. The reality conditions are
$(A^{\pm}_{\bs{k},\lambda_n})^* = A^{\mp}_{-\bs{k},\lambda_n}$, 
$(A^{\pm}_{\bs{k},-\lambda_n})^* 
= A^{\mp}_{-\bs{k},-\lambda_n}$ for $|\bs{k}|^2 > \lambda_n$,
$(A^{\pm}_{\bs{k},-\lambda_n})^* 
= 
A^{\pm}_{-\bs{k},-\lambda_n}$ for $|\bs{k}|^2 < \lambda_n$,
and $ (\epsilon^{\rm L,R }_{{\cal J},n})^*  = \epsilon^{\rm R,L }_{{\cal J},n}$.

Using $\epsilon_{{\cal I},n}^{\rm R} \ri \Lambda_{\cal IJ} \epsilon^{ \rm L}_{{\cal J},m} = -\epsilon_{{\cal I},n}^{\rm L} \ri \Lambda_{\cal IJ} \epsilon^{\rm R}_{{\cal J},m} = \lambda_n \delta_{nm}$ (no summation over $n$) and $\epsilon_{{\cal I},n}^{h} \ri \Lambda_{\cal IJ} \epsilon^{h}_{{\cal J},m} = 0$, we can decompose the integral into the contributions of the positive and negative eigenvalue modes ($\pm\lambda_n$) as
\begin{align}
&
\int \rd^{D-1} \bs{x}
F^2_{IJ}
\bar{E}_{I, i_1...i_{p_I}} 
\der_0 
\delta f_J^{0 i_1...i_{p_I}}
\nonumber \\
&
=
- \sum_{n =1}^{{\cal N}}
\int_{|\bs{k}|^2 < \lambda_n} 
\fr{{\rd}^{D-1} \bs{k}}{(2\pi)^{D-1}}
\lambda_n \sr{ \lambda_n - |\bs{k}|^2 }
\nonumber \\
&
\hph{
\quad =  \sum_{n =1}^{{\cal N}}
\int_{|\bs{k}|^2 < \lambda_n} 
\fr{{\rd}^{D-1} \bs{k}}{(2\pi)^{D-1}}
}
\times 
(|A^{+}_{\bs{k},-\lambda_n}|^2
{\re}^{2 t \sr{\lambda_n - |\bs{k}|^2 }}
- 
|A^{-}_{\bs{k},-\lambda_n}|^2
{\re}^{-2 t \sr{\lambda_n - |\bs{k}|^2 }}
)
\nonumber \\
&
\quad
-2 \sum_{n =1}^{{\cal N}}
\int_{|\bs{k}|^2 > \lambda_n}  \fr{{\rd}^{D-1} \bs{k}}{(2\pi)^{D-1}}
\lambda_n \sr{|\bs{k}|^2 + \lambda_n}
\im \left[
(A^{-}_{\bs{k},-\lambda_n})^*
 A^{+}_{\bs{k},-\lambda_n} 
{\re}^{2{\ri}  t \sr{|\bs{k}|^2 - \lambda_n}}
\right]
\nonumber \\
&
\quad
-
2 \sum_{n =1}^{{\cal N}}
\int \fr{{\rd}^{D-1} \bs{k}}{(2\pi)^{D-1}}
\lambda_n \sr{|\bs{k}|^2 + \lambda_n}
\im \left[
(A^{-}_{\bs{k},\lambda_n})^*
 A^{+}_{\bs{k},\lambda_n} 
{\re}^{2{\ri}  t \sr{|\bs{k}|^2 + \lambda_n}}
\right]\,.
\end{align}
Here, we used 
\begin{equation}
\int \fr{{\rd}^{D-1}\bs{k}}{(2\pi)^{D-1}} 
\lambda_n 
\sr{|\bs{k}|^2 + \lambda_n } \, 
|A^{+}_{\bs{k},\lambda_n}|^2
= 
\int \fr{{\rd}^{D-1}\bs{k}}{(2\pi)^{D-1}} 
\lambda_n 
\sr{|\bs{k}|^2 + \lambda_n } 
 \,
|A^{-}_{\bs{k},\lambda_n}|^2
\end{equation}
for the positive eigenvalue sector.
In the presence of non-zero amplitude of the unstable modes, $|A^{+}_{\bs{k},-\lambda_n}|^2$, the right-hand side becomes negative due to the growth of the unstable mode. 
Therefore, we have
\begin{equation}
    \int \rd^{D-1} \bs{x}
F^2_{IJ}
\bar{E}_{I, i_1...i_{p_I}} 
\der_0 
\delta f_J^{0 i_1...i_{p_I}} < 0.
\end{equation}
In particular, the magnitude of at least one electric field decreases.

\subsection{\label{GMH}Generation of linked configurations}

We can also show that generalized chiral instabilities lead to a gauge field configuration with at least one non-trivial linking number.
We begin with the equation of motion in \er{EOM_aI}, which leads to the conservation of the following quantity:
\begin{equation}
Q_{I}(\Sigma_{D - p_I -1}) 
=
\int_{\Sigma_{D - p_I -1}} 
\(
(-1)^{p_I} F_{IJ}^2 *{\rd}a_J
+
\fr{1}{2}  M_{IJK} a_J \wed {\rd}a_K
\)\,,
\label{Q_generalized}
\end{equation}
where $\Sigma_{D - p_I -1}$ is a $(D - p_I -1)$-dimensional closed subspace. In particular, by taking $\Sigma_{D - p_I -1}$ as a spatial plane perpendicular to the initial field, the conserved charge can be written as
\begin{align}
Q_I (\Sigma_{D - p_I -1}) 
&
=
\int_{\Sigma_{D - p_I -1}} 
\Bigg(
(-1)^{p_I} 
\epsilon_{i_1...i_{D - p_I -1} 0 j_1...j_{p_I}} 
F_{IJ}^2 f_{J}^{0 j_1...j_{p_J}} 
 {\rd}x^{i_1}\weds {\rd}x^{i_{D - p_I -1}}
\nonumber \\
&
\quad
\hph{\int_{D - p_I -1} 
\Bigg(\quad }
+
\fr{1}{2} M_{IJK} a_J \wed {\rd}a_K
\Bigg)\,.
\end{align}
If we take the spatial plane at the initial time slice before the instability, the conserved charge is dominated by the initial electric field $ F_{IJ}^2 \bar{f}_{J}^{0 j_1...j_{p_J}}$. On the other hand, if we take the plane at a time slice after the electric field $F_{IJ}^2 f_{J}^{0j_1...j_{p_I}}$ decreases, the quantity
\begin{equation}
T_I (\Sigma_{D- p_I -1}) 
:= 
\fr{M_{IJK} }{2} \int_{\Sigma_{D- p_I -1}}  a_J \wed {\rd}a_K
\end{equation}
is generated due to the conservation law. This may be regarded as a higher-form generalization of the magnetic helicity in \er{T(V)}, so we call it the generalized magnetic helicity.

Let us show that the generalized magnetic helicity is proportional to the linking of the fluxes of $a_J$ and $a_K$. We assume that the configurations of the dynamical gauge fields are given by closed flux tubes localized on a closed subspace 
$\Sigma_{D - p_I -1}$,
\begin{equation}
{\rd}a_I = m_I \delta_{p_I +1} (\Sigma_{D - p_I -1}),
\label{aI_delta}
\end{equation}
where $\delta_{p_I +1} (\Sigma_{p_I+1})$ is a delta function $(p_I+1)$-form defined by
\begin{equation}
    \int 
    \omega_{D - p_I -1} \wed  
    \delta_{p_I +1} (\Sigma_{D - p_I -1})
     = 
    \int_{\Sigma_{D - p_I -1}} 
    \omega_{D - p_I -1} 
\end{equation}
for a $(D - p_I -1)$-form field $\omega_{D - p_I -1}$. Since $\Sigma_{D-p_I -1}$ is closed, there exists a $(D-p_I)$-dimensional subspace $\Omega_{D - p_I}$ satisfying $\der \Omega_{D - p_I} = \Sigma_{D-p_I -1}$. The quantity $m_I$ is the flux of $a_I$, 
\begin{equation}
 \int_{\Omega'_{p_I +1}} {\rd}a_I = m_I 
    \int_{\Omega'_{p_I +1}}  \delta_{p_I +1} (\Sigma_{D - p_I -1}) = m_I\,,
\end{equation}
where $\Omega'_{p_I +1}$ is a $(p_I+1)$-dimensional subspace intersecting 
$\Sigma_{D-p_I - 1}$ once.

By using the relation obtained by the Stokes theorem, $(-1)^{D - p_I} \rd \delta_{p_I} (\Omega_{D-p_I}) = \delta_{p_I + 1} (\Sigma_{D - p_I -1})$, we can solve \er{aI_delta} in terms of $a_I$ as
\begin{equation}
    a_I = (-1)^{D - p_I} m_I
    \delta_{p_I} (\Omega_{D-p_I}).
\end{equation}
Note that there is a redundancy of the choice of $\Omega_{D-p_I}$, which corresponds to the gauge redundancy. By substituting the solution, we obtain 
\begin{equation}
\begin{split}
T_{I} (\Sigma_{D - p_I - 1}) 
&
=
M_{IJK} 
m_J m_K 
(-1)^{D- p_J}
\int_{\Sigma_{D - p_I -1 }} 
\delta_{p_J} (\Omega_{D-p_J}) 
\wed 
\delta_{p_K+1} (\Sigma_{D-p_K-1})
\\
&
=
M_{IJK} 
m_J m_K 
(-1)^{D- p_J}
\link (\Sigma_{D-p_J -1}, \Sigma_{D-p_K-1} ; 
\Sigma_{D - p_I  - 1}),
\end{split}
\end{equation}
where $\link (\Sigma_{D-p_J -1}, \Sigma_{D-p_K-1}; \Sigma_{D - p_I  - 1})$ is the linking number of $\Sigma_{D-p_J -1}$ and $\Sigma_{D-p_K-1}$ on $\Sigma_{D - p_I -1}$.

\subsection{Non-invertible symmetries}

We have obtained the conserved charge including the generalized magnetic helicity in \er{Q_generalized}. This conserved charge is again not invariant under the large gauge transformations due to the existence of the topological term $\int_{\Sigma_{D-p_I-1}}\fr{1}{2}  M_{IJK} a_J \wed {\rd}a_K$. If we try to construct an invertible symmetry generator associated with  the conserved charge $\re^{\ri\alpha  Q_I (\Sigma_{D - p_I -1})}$ with a real parameter $\alpha$, the parameter $\alpha$ should be constrained as $\re^{\ri\alpha} = 1$ as discussed in section~\ref{CPInoninv}. 
Instead, we can construct a non-invertible symmetry generator by modifying the topological term. As the construction of non-invertible symmetries for generic $M_{IJK}$ will depend on the detailed structure of $M_{IJK}$ (e.g., the existence of the inverse of the matrix $(M_{IJ})_K$ for a fixed $K$) and will be complicated, we here consider some specific cases of $M_{IJK}$ without restricting the rank of the $p$-form fields.

\subsubsection{Case 1: one gauge field}

Let us first consider the case with one $p_1$-form field $a_1$ in $D$-dimensional spacetime. The action is given by
\begin{equation}
 S_1 =  \int \(
- \fr{F_1^2}{2} |{\rd}a_1|^2 
+ 
\fr{1}{3! \cdot 4\pi^2} a_1 \wed {\rd}a_1 \wed \rd a_1
\)\,,
\end{equation}
where $p_1$ is an odd integer satisfying $3p_1 + 2 = D$, such as $(D,p_1)= (5,1), (11,3), ...$. The gauge field is normalized by the flux quantization condition as $\int_{\Sigma_{p_1 +1}} {\rd}a_1 \in 2\pi \bb{Z}$. The case $(D,p_1)= (5,1)$ is called Maxwell-Chern-Simons theory, and the case $(D,p_1)= (11,3)$ appears in the 11-dimensional supergravity. The non-invertible symmetries in those theories have been recently studied in refs.~\cite{Damia:2022bcd,Garcia-Valdecasas:2023mis}.

We now see the non-invertible symmetry generator from the equation of motion, 
\begin{equation}
 (-1)^{p_1}
 F_1^2 {\rd} *{\rd} a_1 
+ \fr{1}{2! \cdot 4\pi^2}  {\rd}a_1 \wed {\rd}a_1
 =0\,.
\end{equation}
By the equation of motion, we have the symmetry generator on a $(D - p_1 -1 )$-dimensional closed subspace $\Sigma_{D-p_1 - 1 }$,
\begin{equation}
 \exp \( \ri \alpha_1 \int_{\Sigma_{D-p_1 - 1 }} \left[ (-1)^{p_1} 
F_1^2 *{\rd} a_1 + \fr{1}{2! \cdot 4\pi^2} a_1 \wed {\rd}a_1 \right] \)\,,
\end{equation}
but it is not large gauge invariant due to the gauge non-invariant topological term $\exp\( \ri \alpha_1 \int_{\Sigma_{D-p_1 - 1 }} \fr{1}{2! \cdot 4\pi^2} a_1 \wed {\rd}a_1\)$. For $\alpha_1 = \fr{2\pi}{q_1}$ with $q_1 \in \bb{Z}$, we can modify the topological term in terms of the path integral of the topological field theory. The modification can be done as follows:
\begin{equation}
  \exp\(\fr{\ri}{4\pi q_1} 
\int_{\Sigma_{D-p_1 - 1 }} a_1 \wed {\rd}a_1\)
\to 
\int {\cal D}c_1
  \exp\(\fr{\ri}{4\pi} 
\int_{\Sigma_{D-p_1 - 1 }} (- q_1 c_1 \wed {\rd}c_1 + 2 c_1 \wed {\rd}a_1)\)\,,
\end{equation}
where $c_1$ is a $p_1$-form gauge field living on $\Sigma_{D-p_1 - 1 }$. Here and below, we assume that subspaces for the symmetry defects do not have self-intersections. The normalization of $c_1$ is given by $\int_{\Sigma_{p_1 +1}} \rd c_1 \in 2\pi \bb{Z}$. We can thus introduce the following non-invertible symmetry defect:
\begin{equation}
 D_1 (\re^{\fr{2\pi \ri}{q_1}}, \Sigma_{D - p_1 -1})
 = 
 \re^{\fr{2\pi \ri }{q_1} 
\int_{\Sigma_{D-p_1 - 1 }} (-1)^{p_1} 
F_1^2 *{\rd} a_1 }
\int {\cal D}c_1
\re^{\fr{ \ri }{4\pi} 
\int_{\Sigma_{D-p_1 - 1 }} (- q_1 c_1 \wed {\rd}c_1 + 2 c_1 \wed {\rd}a_1)}\,. 
\end{equation}

As in section~\ref{CPInoninv}, the symmetry defect can be understood as a generalization of the worldsheet of a magnetic flux tube. Let us construct a linked configuration of the magnetic flux tubes from an unlinked one in terms of the non-invertible symmetry defects as follows.
First, we consider the correlation function $\vevs{ D_1 (\re^{\fr{2\pi \ri}{q_1}}, \Sigma^0_{D - p_1 -1}) D_1 (\re^{\fr{2\pi \ri}{q''_1}}, \Sigma''_{D - p_1 -1})}$, where $\Sigma^0_{D - p_1 -1}$ and $\Sigma''_{D - p_1 -1}$ are not intersected, and they can be shrunk without intersections. We then intersect these two objects by a continuous deformation of $\Sigma^0_{D - p_1 -1}$ to $\Sigma^1_{D - p_1 -1}$. We then have
\begin{align}
& \vevs{ D_1 (\re^{\fr{2\pi \ri}{q_1}}, \Sigma^0_{D - p_1 -1})
 D_1 (\re^{\fr{2\pi \ri}{q''_1}}, \Sigma''_{D - p_1 -1})}
\nonumber \\
&
=
   \vevs{
   \re^{\fr{\ri}{q_1 q_1''} 
\int {\rd}a_1 \wed 
\delta_{p_1}(\Omega''_{D - p_1} ) 
\wed
 \delta_{p_1 +1}(\Sigma^1_{D - p_1 -1} )
 }
 D_1 (\re^{\fr{2\pi \ri}{q_1}}, \Sigma^1_{D - p_1 -1})
 D_1 (\re^{\fr{2\pi \ri}{q''_1}}, \Sigma''_{D - p_1 -1})
},
\end{align}
where $\Omega''_{D - p_1} $ is a $(D-p_1)$-dimensional subspace whose boundary is $\Sigma''_{D-p_1-1}$.

We now relate the generalized magnetic helicity to the linking number.
This can be done by further inserting the non-invertible symmetry defect to the correlation function as $\vevs{ D_1 (\re^{\fr{2\pi \ri}{q_1}}, \Sigma^0_{D - p_1 -1}) D_1 (\re^{\fr{2\pi \ri}{q'_1}}, \Sigma^{\prime 0}_{D - p_1 -1}) D_1 (\re^{\fr{2\pi \ri}{q''_1}}, \Sigma''_{D - p_1 -1})}$, where $\Sigma^0_{D - p_1 -1}$, $\Sigma^{\prime 0}_{D - p_1 -1}$ and $\Sigma''_{D - p_1 -1}$ are not intersected each other.
We can intersect them by continuously deforming $\Sigma^0_{D - p_1 -1}$ and $\Sigma^{\prime 0}_{D - p_1 -1}$ to $\Sigma^{1}_{D - p_1 -1}$ and $\Sigma^{\prime 1}_{D - p_1 -1}$, respectively, and obtain the following correlation function:
\begin{align}
&
   \vevs{ D_1 (\re^{\fr{2\pi \ri}{q_1}}, \Sigma^0_{D - p_1 -1})
 D_1 (\re^{\fr{2\pi \ri}{q'_1}}, \Sigma^{\prime 0}_{D - p_1 -1})
D_1 (\re^{\fr{2\pi \ri}{q''_1}}, \Sigma''_{D - p_1 -1})
}
\nonumber \\
&
=
\re^{\fr{2 \pi \ri}{q_1 q_1' q_1''} 
\int \delta_{p_1 +1} (\Sigma^1_{D - p_1 -1})  
\wed
\delta_{p_1}(\Omega''_{D - p_1} ) 
\wed
 \delta_{p_1 +1}(\Sigma^{\prime 1}_{D - p_1 -1} )
 }
\nonumber \\
&
\qquad
\times 
\langle
\re^{\fr{\ri}{q'_1 q_1''} 
\int {\rd}a_1 \wed 
\delta_{p_1} (\Omega''_{D - p_1} ) 
\wed
 \delta_{p_1 +1}(\Sigma^{\prime 1}_{D - p_1 -1} )
 }
 \re^{\fr{\ri}{q_1 q_1'} 
\int {\rd}a_1 \wed 
\delta_{p_1}(\Omega^{1}_{D - p_1} ) 
\wed
 \delta_{p_1 +1}(\Sigma^{\prime 1}_{D - p_1 -1} )
 }
\nonumber \\
&
\qquad\quad
\times 
\re^{\fr{\ri}{q_1 q_1''} 
\int {\rd}a_1 \wed 
\delta_{p_1}(\Omega''_{D - p_1} ) 
\wed
 \delta_{p_1 +1}(\Sigma^1_{D - p_1 -1} )
 }
\nonumber \\
&
\qquad\quad
\times 
 D_1 (\re^{\fr{2\pi \ri}{q_1}}, \Sigma^1_{D - p_1 -1})
 D_1 (\re^{\fr{2\pi \ri}{q'_1}}, \Sigma'_{D - p_1 -1})
D_1 (\re^{\fr{2\pi \ri}{q''_1}}, \Sigma''_{D - p_1 -1})
\rangle\,.
\end{align}
Here, $\Omega^1_{D-p_1}$ is a $(D-p_1)$-dimensional subspace whose boundary is $\Sigma^1_{D-p_1-1}$. The integral
$\int \delta_{p_1 +1} (\Sigma^1_{D - p_1 -1})  
\wed
\delta_{p_1}(\Omega''_{D - p_1})
\wed
 \delta_{p_1 +1}(\Sigma^{\prime 1}_{D - p_1 -1})
 $
can be understood as a linking number between $\Sigma^{1}_{D - p_1}$ and $\Sigma''_{D - p_1 -1}$ on $\Sigma^{\prime 1}_{D - p_1-1}$. By taking $\Sigma^{1}_{D - p_1}$ and $\Sigma''_{D - p_1 -1}$ as temporally extended subspaces and $\Sigma^{\prime 1}_{D - p_1-1}$ as a temporally localized one, this correlation function means that the symmetry defect $D_1 (\re^{\fr{2\pi \ri}{q_1'}}, \Sigma^{\prime 1}_{D - p_1-1})$ containing a gauge-invariant formulation of the generalized magnetic helicity $\fr{1}{2!\cdot 4\pi^2} \int_{\Sigma^{\prime 1}_{D - p_1 - 1}} a_1 \wed \rd a_1$ captures the linking number of $\Sigma^{1}_{D - p_1}$ and $\Sigma''_{D - p_1 -1}$ on $\Sigma^{\prime 1}_{D - p_1-1}$ proportional to the magnetic fluxes $\fr{2\pi}{q_1}$ and $\fr{2\pi}{q''_1}$.

\subsubsection{Case 2: two gauge fields}

We then consider a system with $p_2$- and $p_3$-form gauge fields $a_2$ and $a_3$, whose action is given by
\begin{equation}
 S_2 = \int \(
 -\fr{F_2^2}{2} |{\rd}a_2|^2 
- \fr{F_3^2}{2} |{\rd}a_3|^2 
+ \fr{1}{2! \cdot 4\pi^2}
a_3 \wed {\rd}a_2 \wed {\rd}a_2 
\
\).
\end{equation}
Here, we assume that $p_2$ and $p_3$ satisfy $2p_2 + p_3 +2 = D$ and $p_2$ is an odd integer. The normalizations of the gauge fields are given by the flux quantization conditions, $\int_{\Sigma_{p_2 +1}} {\rd} a_2$, $\int_{\Sigma_{p_3 +1}} {\rd} a_3 \in 2\pi \bb{Z}$.

This case includes the axion electrodynamics, where the dimension and the ranks of the gauge fields are $(D, p_2, p_3) = (4,1,0)$, i.e., $a_2$ and $a_3$ correspond to the gauge field $a$ and the axion field $\phi$, respectively.

Since the derivation of the existence of non-invertible symmetries is the same as Case 1, we simply display the results. The equations of motion for $a_2$ and $a_3$ are given by
\begin{gather}
(-1)^{p_2}
F_2^2 \rd* {\rd}a_2 
+
\fr{1}{ 4\pi^2} {\rd}a_2 \wed {\rd} a_3 =0\,,
\\
(-1)^{p_3} 
F_3^2 \rd * {\rd}a_3 
+ \fr{1}{2! \cdot 4\pi^2} {\rd}a_2 \wed {\rd} a_2 =0\,.
\end{gather}
By the equations of motion, we can construct non-invertible symmetry defects as
\begin{align}
 D_2 (\re^{\fr{2\pi \ri}{q_2}}, \Sigma_{D - p_2 -1})
&
= 
 \re^{\fr{2\pi \ri }{q_2} 
\int_{\Sigma_{D-p_2 - 1 }} (-1)^{p_2} 
F_2^2 * {\rd} a_2 }
\nonumber
\\
&
\quad
\times 
\int {\cal D}[c_2, c_2'] 
\re^{\fr{\ri}{2\pi} 
\int_{\Sigma_{D-p_3 - 1 }} (- q_2 c_2 \wed {\rd}c'_2 
+ c_2 \wed {\rd}a_3 + 
a_2 \wed {\rd} c_2' )}\,,
\\
 D_3 (\re^{\fr{2\pi \ri}{q_3}}, \Sigma_{D - p_3 -1})
 &= 
 \re^{\fr{2\pi \ri }{q_3} 
\int_{\Sigma_{D-p_3 - 1 }} (-1)^{p_3} 
F_3^2 * {\rd} a_3 }
\nonumber 
\\
&
\quad
\times \int {\cal D}c_3 \re^{\fr{\ri}{4\pi} 
\int_{\Sigma_{D-p_4 - 1 }} (- q_3 c_3 \wed {\rd}c_3 + 2 c_3 \wed {\rd}a_2)}\,. 
\end{align}
Here, $c_2$, $c_2'$, and $c_3$ are $p_2$-, $p_3$-, $p_2$-form $U(1)$ gauge fields, respectively.

The relation between the generalized magnetic helicity and the linking number can be found by the correlation function as
\begin{align}
 &
 \vevs{
    D_2 (\re^{\fr{2\pi \ri}{q_2}}, \Sigma^0_{D - p_2 -1})
     D_2 (\re^{\fr{2\pi \ri}{q_2}}, \Sigma^{\prime 0}_{D - p_2 -1})
      D_3 (\re^{\fr{2\pi \ri}{q_3}}, \Sigma_{D - p_3 -1})
    }
    \nn
    \\
    &
    =
    \re^{\fr{2\pi \ri }{q_2 q_2' q_3} \int 
    \delta_{p_2 +1} (\Sigma^{1}_{D-p_2- 1})
    \wed
    \delta_{p_2 +1} (\Sigma^{\prime 1}_{D-p_2- 1})
    \wed
    \delta_{p_3 } (\Omega_{D - p_3})
    }
    \nn
    \\
    &
    \quad
    \times
    \langle
        \re^{\fr{\ri }{q_2 q_2'} 
    \int_{\Omega_{D - p_3}
    } \rd a_2 \wed \delta_{p_2 +1} (\Sigma_{D-p_2-1}^{1})
    }
            \re^{\fr{\ri }{q_2 q_3} 
    \int_{\Omega^{\prime 1}_{D - p_2}
    } \rd a_3 \wed \delta_{p_2 +1} (\Sigma_{D-p_2-1}^{1})
    }
    \nn
    \\
    &
    \qquad\quad 
    \times
    \re^{\fr{\ri }{q_2' q_3} 
    \int_{\Omega_{D - p_3}
    } \rd a_2 \wed \delta_{p_2 +1} (\Sigma_{D-p_2-1}^{\prime 1})
    }
    \nn 
    \\
    &
    \qquad\quad 
    \times 
    D_2 (\re^{\fr{2\pi \ri}{q_2}}, \Sigma^1_{D - p_2 -1})
     D_2 (\re^{\fr{2\pi \ri}{q_2}}, \Sigma^{\prime 1}_{D - p_2 -1})
      D_3 (\re^{\fr{2\pi \ri}{q_3}}, \Sigma_{D - p_3 -1})
    \rangle\,.
\end{align}
The symmetry defect $ D_2 (\re^{\fr{2\pi \ri}{q_2}}, \Sigma^1_{D - p_2 -1})$ detects the generalized magnetic helicity $\fr{1}{4 \pi^2} \int_{\Sigma^1_{D - p_1 -1} }a_2 \wed \rd a_3$ for the configuration $ D_2 (\re^{\fr{2\pi \ri}{q_2}}, \Sigma^{\prime 1}_{D - p_2 -1}) D_3 (\re^{\fr{2\pi \ri}{q_3}}, \Sigma_{D - p_3 -1})$ as a phase factor.
For instance, we take $\Sigma^{\prime 1}_{D-p_2- 1}$ and $\Sigma_{D - p_3 - 1}$ as temporally extended subspaces and $\Sigma^{1}_{D-p_2- 1}$ as a temporally localized subspace, the integral means that the linking number of the magnetic fluxes on $\Sigma^{\prime 1}_{D-p_2- 1} \cap \Sigma^{1}_{D-p_2- 1}$ and $\Sigma_{D - p_3 - 1} \cap \Sigma^{1}_{D-p_2- 1}$.

\subsubsection{Case 3: three gauge fields}

Finally, we consider the case with the $p_4$-, $p_5$-, and $p_6$-form fields $a_4$, $a_5$, and $a_6$ in $D$-dimensional spacetime.
The action is given by
\begin{equation}
 S_3 = \int \(
 -\fr{F_4^2}{2} |{\rd}a_4|^2 
- \fr{F_5^2}{2} |{\rd}a_5|^2 
- \fr{F_6^2}{2} |{\rd}a_6|^2
+
\fr{1}{4\pi^2} a_4 \wed {\rd}a_5 \wed {\rd} a_6
\)\,,
\end{equation}
where the ranks of the gauge fields satisfy $p_4 +p_5 +p_6 +2 = D$. The gauge fields are normalized by the flux quantization conditions as 
$\int_{\Sigma_{p_4 +1}} {\rd} a_4$,
$\int_{\Sigma_{p_5 +1}} {\rd} a_5$, 
$\int_{\Sigma_{p_6 +1}} {\rd}a_6 \in 2\pi \bb{Z}$.
We will consider an example $(D, p_4,p_5,p_6)= (3,1,0,0)$ in section~\ref{3d}.

From the equations of motion for $a_{4,5,6}$, 
\begin{align}
 (-1)^{p_4}
 F_4^2 {\rd} * {\rd} a_4 
 + 
\fr{1}{4\pi^2} {\rd} a_5 \wed {\rd}a_6  &=0\,,
\\
(-1)^{p_5 + D (p_4+1)}
F_5^2 {\rd} * {\rd} a_5 
 + 
\fr{1}{4\pi^2}
\rd a_6 \wed {\rd} a_4   &=0\,,
\\
(-1)^{p_6 + D (p_6+1)} 
F_6^2 {\rd} * {\rd} a_6 
 + 
\fr{1}{4\pi^2}
\rd a_4 \wed {\rd}a_5  &=0\,,
\end{align}
the non-invertible symmetry defects are obtained as 
\begin{align}
 D_4 (\re^{\fr{2\pi \ri}{q_4}}, \Sigma_{D - p_4 -1})
&
= 
 \re^{\fr{2\pi \ri }{q_4} 
\int_{\Sigma_{D-p_4 - 1 }} (-1)^{p_4} 
F_4^2 * {\rd} a_4 }
\nonumber
\\
&
\quad
\times 
\int {\cal D}[c_4, c_4'] 
\re^{\fr{\ri}{2\pi} 
\int_{\Sigma_{D-p_4 - 1 }} (- q_4 c_4 \wed {\rd}c'_4 
+ a_5 \wed {\rd} c'_4 
+ c_4 \wed {\rd} a_6)}\,,
\\
 D_5 (\re^{\fr{2\pi \ri}{q_5}}, \Sigma_{D - p_5 -1})
&
= 
 \re^{\fr{2\pi \ri }{q_5} 
\int_{\Sigma_{D-p_5 - 1 }} (-1)^{p_5 + D(p_4 +1)} 
F_4^2 * {\rd} a_5 }
\nonumber
\\
&
\quad
\times 
\int {\cal D}[c_5, c_5'] 
\re^{\fr{\ri}{2\pi} 
\int_{\Sigma_{D-p_5 - 1 }} (- q_5 c_5 \wed {\rd} c'_5 
+ a_6 \wed {\rd} c'_5 + 
c_5 \wed {\rd} a_4)}\,,
\\ 
D_6 (\re^{\fr{2\pi \ri}{q_6}}, \Sigma_{D - p_6 -1})
&
= 
 \re^{\fr{2\pi \ri }{q_6} 
\int_{\Sigma_{D-p_6 - 1 }} (-1)^{p_6 + D (p_6+1)} 
F_6^2 * {\rd} a_6 }
\nonumber
\\
&
\quad
\times 
\int {\cal D}[c_6, c_6'] 
\re^{\fr{\ri}{2\pi} 
\int_{\Sigma_{D-p_6 - 1 }}
(- q_6 c_6 \wed {\rd}c'_6 
+ a_4 \wed {\rd} c'_6
+ c_6 \wed {\rd} a_5 )}\,.
\end{align}
Here, the ranks of the $U(1)$ gauge fields $c_4$, $c_4'$, $c_5$, $c_5'$, $c_6$, and $c_6'$ are $p_5$, $p_6$, $p_6$, $p_4$, $p_4$, and $p_5$, respectively. 

The relation between the generalized magnetic helicity and the linking number can be obtained by the same procedure as the previous cases:
\begin{align}
 &
 \vevs{
    D_4 (\re^{\fr{2\pi \ri}{q_4}}, \Sigma^0_{D - p_4 -1})
     D_5 (\re^{\fr{2\pi \ri}{q_5}}, \Sigma^{0}_{D - p_5 -1})
      D_6 (\re^{\fr{2\pi \ri}{q_6}}, \Sigma_{D - p_6 -1})
    }
    \nn
    \\
    &
    =
    \re^{ (-1)^{D (p_5+1)}
    \fr{2\pi \ri }{q_4 q_5 q_6} \int 
    \delta_{p_4 +1} (\Sigma^{1}_{D-p_4- 1})
    \wed
    \delta_{p_5 +1} (\Sigma^{1}_{D-p_5 - 1})
    \wed
    \delta_{p_6 } (\Omega_{D - p_6})
    }
    \nn
    \\
    &
    \quad
    \times
    \langle
        \re^{(-1)^{(p_4+1)(p_6+1)} 
        \fr{\ri }{q_4 q_6} 
    \int_{\Omega_{D - p_6}
    } \rd a_5 \wed \delta_{p_4 +1} (\Sigma_{D-p_4-1}^{1})
    }
    \nn
    \\
    &
    \qquad\quad
    \times
            \re^{(-1)^{D (p_4+1)} 
            \fr{\ri }{q_4 q_5} 
    \int_{\Omega^{ 1}_{D - p_5}
    } \rd a_6 \wed \delta_{p_4 +1} (\Sigma_{D-p_4-1}^{1})
    }
    \nn
    \\
    &
    \qquad\quad 
    \times
    \re^{(-1)^{D (p_6+1)} \fr{\ri }{q_5 q_6} 
    \int_{\Omega_{D - p_6}
    } \rd a_4 \wed \delta_{p_5 +1} (\Sigma_{D-p_5-1}^{ 1})
    }
    \nn 
    \\
    &
    \qquad\quad 
    \times 
    D_4 (\re^{\fr{2\pi \ri}{q_4}}, \Sigma^1_{D - p_4 -1})
     D_5 (\re^{\fr{2\pi \ri}{q_5}}, \Sigma^{ 1}_{D - p_5 -1})
      D_6 (\re^{\fr{2\pi \ri}{q_6}}, \Sigma_{D - p_6 -1})
    \rangle\,.
\end{align}

\section{\label{appli}Applications}

In this section, we first apply the mechanism of the generalized chiral instabilities to the instability of the $(3+1)$-dimensional axion electrodynamics in electric field~\cite{Bergman:2011rf,Ooguri:2011aa}. While the existence of the instabilities itself is known, our new finding is the non-trivial linking numbers of the resulting configurations characterized by the generalized magnetic helicity (which we will show without assuming specific boundary conditions). We also show that this mechanism is realized in $(2+1)$-dimensional Goldstone-Maxwell model in electric field.

\subsection{\label{sec:axinst}Axion electrodynamics in electric field}

We here consider the instability of the $(3+1)$-dimensional axion electrodynamics in the homogeneous electric field. 
The action is given by
\begin{equation}
S = \int \( -\fr{v^2}{2} |\rd\phi|^2 
-
\fr{1}{2e^2} |\rd a|^2  
+\fr{1}{8\pi^2} 
\phi {\rd}a \wed \rd a\)\, ,
\end{equation}
which is the same as the action in  \er{CPIaction}. Note that the case of the massive axion is considered in refs.~\cite{Bergman:2011rf,Ooguri:2011aa}. In this case, the finite mass of the axion field opens up an infrared region where the instability does not occur. On the other hand, we here consider the massless axion for simplicity. As we will see below, the instability occurs even in the infrared limit.

Let us check the existence of the instability. The equations of motion are 
\begin{gather}
    v^2 \rd*\rd \phi 
    +
    \fr{1}{4\pi^2} {\rd}a \wed \rd a =0\,,
    \\
    - \fr{1}{e^2} \rd* {\rd}a 
    +
    \fr{1}{4\pi^2} \rd\phi \wed \rd a =0\,.
    \label{EOM_a2}
\end{gather}
We assume the initial constant electric field of the form
\begin{equation}
 \rd \bar{a}   =  \bar{f}_{0i} {\rd}x^0 \wed {\rd}x^i 
=- \bar E_i {\rd}x^0 \wed {\rd}x^i.
\end{equation}
In this case, there is an instability for a linear combination of $\phi$ and $a$. We decompose the 1-form gauge field $a$ into the initial configuration $\bar{a}$ and the fluctuation $\delta a$ as $a  = \bar{a} + \delta a$. The equations of motion for $\phi$ and $\delta a$ can be written in momentum space as
\begin{equation}
 (\omega^2 - |\bs{k}|^2) \hat{a}_{\cal I}
= \ri k_l \hat{M}^l_{\cal IJ} \hat{a}_{\cal J}.
\end{equation}
Here, we took the temporal gauge $a_0 =0$ and defined $\hat{a}_{\cal I} = (\hat{\phi}, \hat{a}_1, \hat{a}_2 , \hat{a}_3)$ as Fourier transformations of $(v\phi, \fr{1}{e} \delta a_1, \fr{1}{e} \delta  a_2,\fr{1}{e} \delta  a_3)$.
The matrix characterizing the instability can be derived as 
\begin{equation}
\hat{M}^{l}_{\cal IJ} k_l 
=
- \fr{e}{4\pi^2 v}
 \mtx{
 0 & 
 \epsilon^{l 1 0 k}\bar E_k k_l 
 &
 \epsilon^{l 2 0 k}\bar E_k k_l 
 & 
 \epsilon^{l 3 0 k}\bar E_k k_l
\\
\epsilon^{1  l0 k}\bar E_k k_l
\\
\epsilon^{2  l0 k}\bar E_k k_l
\\
\epsilon^{3 l 0k}\bar E_k k_l
}\,.
\end{equation}
The matrix $ \hat{M}^l_{\cal IJ}k_l$ can be block diagonalized by using an orthogonal matrix $P_{\cal IJ}$ as 
\begin{equation}
P_{\cal K I} \hat{M}^l_{\cal K L} k_l P_{\cal L J}
= \mtx{
0 & \lambda 
\\
-\lambda & 0 
\\
&& 0_{2\times 2}
}=: \Lambda_{\cal IJ}\,,
\end{equation}
where 
\begin{equation}
\lambda = \fr{e}{4\pi^2 v} 
\sr{(\bar E_2 k_3- \bar E_3 k_2)^2
+ (\bar E_3 k_1- \bar E_1 k_3)^2
+ (\bar E_1 k_2- \bar E_2 k_1)^2 }\,.
\end{equation}
The dispersion relation is obtained as
\begin{equation}
\omega^2 = |\bs{k}|^2, \quad \omega^2 = |\bs{k}|^2 \pm \lambda.
\end{equation}
Therefore, there is one unstable mode in the infrared region, $0 < |\bs{k}|^2 < \lambda$.

From the argument for the generalized chiral instability in section~\ref{GCI}, we can show the reduction of the initial electric field and the generation of the configuration with linking number. 
Rather than presenting the details of the results specific to the present case, we here provide a physical argument for the existence of the instability, decrease of the initial electric field, and generation of the configuration with the linking number.

\begin{figure}[t]
\centering 
 \ig[width=37.5em]{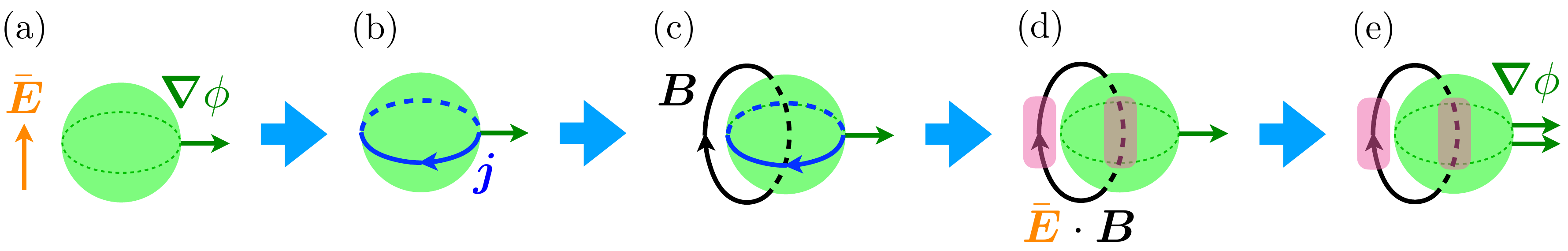}
\caption{\label{axinst} 
Instability in axion electrodynamics in electric field. The axionic domain wall is expressed as the green sphere. 
The value of $\bar{\bs{E}} \cdot \bs{B}$ is non-zero in the regions colored by pink.}
\end{figure}

First, we consider the instability in the homogeneous background electric field. We assume an infinitesimal fluctuation of the axion whose configuration is a spherical domain wall (figure~\ref{axinst}(a)).
The gradient of the axion field, $\bs{\na}\phi$, induces a current $\bs{j} = - C \bar{\bs{E}} \times\bs{\na}\phi$ on the domain wall (figure~\ref{axinst}(b)). This is the anomalous Hall effect~\cite{Sikivie:1984yz,Wilczek:1987mv,Qi:2008ew,Essin:2008rq}. The Amp\`ere law $\bs{\na}\times \bs{B} = \bs{j}$ implies that the induced current gives rise to a magnetic field through the domain wall (figure~\ref{axinst}(c)).
The magnetic field with the background electric field causes $\bar{\bs{E}} \cdot \bs{B}<0$ and $\bar{\bs{E}} \cdot \bs{B}>0$ inside and outside of the sphere, respectively (figure~\ref{axinst}(d)). 
By the equation of motion for the axion $v^2 \bs{\na}\cdot \bs{\na}\phi = - \frac{1}{4\pi^2} \bar{\bs{E}} \cdot \bs{B}$, we see that $\bs{\na}\phi$ increases inside of the sphere, while it decreases outside (figure~\ref{axinst}(e)).
Therefore, the infinitesimal fluctuation of the axion field causes the generation of the magnetic field and the growth of $\bs{\na}\phi$.

\begin{figure}[t]
\vspace{1em}
\centering
 \ig[height=6em]{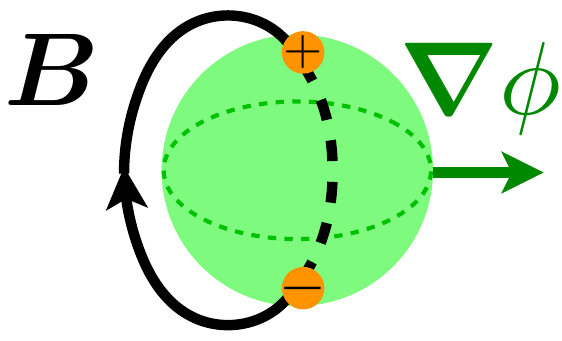} 
\caption{\label{axi-mag} 
Decrease of the initial electric field due to the induced electric charge $- C \bs{B} \cdot \bs{\na} \phi$. The orange dots with the signs represent the induced electric charges. The direction of the induced electric field on the domain wall is opposite to that of the initial electric field.}
\end{figure}

Let us then see the reduction of the initial electric field. The configuration of $\bs{\na }\phi$ and $\bs{B}$ generated by the generalized chiral instability leads to the induced electric charge via the Gauss law $\bs{\na}\cdot \bs{E} = - C \bs{\na}\phi \cdot \bs{B}$. Figure~\ref{axi-mag} shows that the direction of the dielectric polarization by the induced charge is anti-parallel to the direction of the initial electric field. Therefore, the initial electric field tends to be reduced.

\begin{figure}[t]
\centering
\includegraphics[height=5.5em]{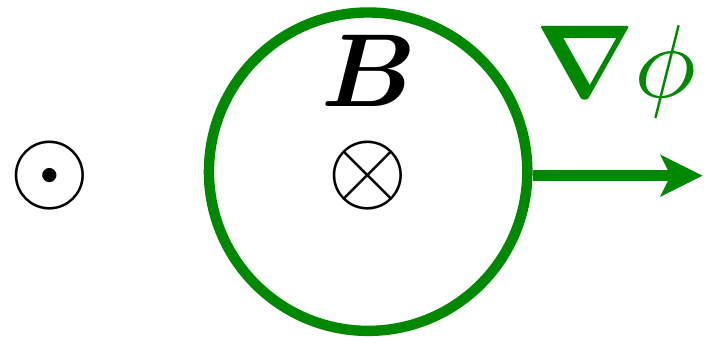}
\caption{\label{axinstlink}
The linked configuration of the axionic domain wall and magnetic field on the surface $S$. The surface is taken so that it contains the equator of the sphere of the thin domain wall. The linking on the surface means that the axion domain wall surrounds either of the magnetic fluxes. 
}
\end{figure}

Finally, we discuss that the resulting configuration is characterized by a non-trivial linking number. From the equation of motion in \er{EOM_a2}, we have a conserved charge, 
\begin{equation}
    Q_a (S)
    = \int_{S} 
    \(-\fr{1}{e^2} \bs{E} \cdot \rd \bs{S}
    + \fr{1}{4\pi^2} \phi \, \rd a \)\,,
\end{equation}
where $S$ is a spatial plane perpendicular to the initial electric field and $\rd S_i = \fr{\epsilon_{i jk}}{2!} \rd x^j \wed \rd x^k $ is a surface element. On the one hand, if we choose the surface $S$ at the initial time, the conserved charge is dominated by the initial electric field. On the other hand, if we choose $S$ at the time after the electric field decreases, the second term 
\begin{equation}
    T_{a} (S) = \fr{1}{4\pi^2} \int_S
    \phi \, {\rd} a
\end{equation}
is generated.

We can also relate $T_{a}(S)$ to a linking number. We consider the configuration of the axion field as a thin domain wall as in figure~\ref{axinstlink}:
\begin{equation}
    \rd \phi = 
    m_\phi \delta_1 ({\cal V}),
    \label{dphi}
\end{equation}
where ${\cal V}$ is a 3-dimensional closed subspace corresponding the location of the worldvolume of the domain wall, and $m_\phi$ is the amplitude of the domain wall. We further consider a magnetic flux tube induced by the anomalous Hall effect,
\begin{equation}
    {\rd}a = m_a \delta_2 ({\cal S}_a), 
\end{equation}
where ${\cal S}_a$ is a 2-dimensional closed subspace corresponding to the location of the worldsheet of the magnetic flux tube. As \er{dphi} can be solved as
\begin{equation}
    \phi = m_\phi \delta_0 (\Omega_{\cal V})
\end{equation}
with a 4-dimensional subspace $\Omega_{\cal V}$ whose boundary is ${\cal V} = \der \Omega_{\cal V}$, we have
\begin{equation}
    T_{a } (S)
    =\fr{m_{\phi } m_a}{4\pi^2} 
    \int_{S} 
    \delta_0 (\Omega_{\cal V}) 
    \delta_2 ({\cal S}_a)
     = \fr{m_{\phi } m_a}{4\pi^2} 
     \link ({\cal V}, {\cal S}_a; S)\,.
\end{equation}
Here, $\link ({\cal V}, {\cal S}_a; S)$ is a linking number between ${\cal V} \cap S$ (it is topologically a circle) and ${\cal S}_a \cap S$ (it is topologically two points) on $S$. Figure~\ref{axinstlink} shows the configuration of the magnetic flux and the domain wall on the integration surface.

We comment on the difference between our setup and ref.~\cite{Ooguri:2011aa}. In the latter, a specific boundary condition is considered with a possible realization in realistic materials in mind. The magnetic field and axion field configurations generated by the instability depend on the choice of the boundary condition for the axion field, corresponding to the interface of the axionic material and a normal insulator. In our case, on the other hand, the gradient of the axion field is generated by the instability independent of the specific boundary condition.

\subsection{\label{3d}$(2+1)$-dimensional Goldstone-Maxwell model in electric field}

Finally, we study the instability for $(2+1)$-dimensional analogue of the Goldstone-Maxwell model~\cite{Cordova:2018cvg} in a homogeneous electric field~\cite{Yamamoto:2022vrh}.

We consider the action~\cite{Yamamoto:2022vrh,Damia:2022rxw}
\begin{equation}
 S_{\phi \chi a} = \int \(
 - \fr{v_\phi^2}{2}|\rd\phi|^2 
 -
 \fr{v_\chi^2}{2}|\rd\chi|^2 
 -
 \fr{1}{2e^2} 
 |\rd a|^2 
 +
 \fr{1}{4\pi^2} a \wed \rd\phi \wed \rd\chi \)\,,
\end{equation}
where $\phi$ and $\chi$ are $2\pi$ periodic scalar fields and $a$ is the gauge field. The equations of motion for $\phi$, $\chi$, and $a$ are
\begin{equation}
\begin{split}
v_\phi^2 \der_\mu \der^\mu  \phi +
 \fr{1}{2! \cdot 4\pi^2} 
 \epsilon^{\mu \nu \rho} \der_\nu \chi 
  f_{\mu \rho} &= 0\,,
\\
v_\chi^2 \der_\mu \der^\mu  \chi -
 \fr{1}{2! \cdot 4\pi^2} 
 \epsilon^{\mu \nu \rho} \der_\nu \phi 
  f_{\mu \rho} &= 0\,,
  \\
  \fr{1}{e^2} \der_\mu f^{\mu \rho }
  - \fr{1}{4\pi^2} 
  \epsilon^{\mu \nu \rho} \der_\mu \phi 
  \der_\nu \chi & =0\,.
    \end{split}
\end{equation}
In the presence of the homogeneous initial electric field $\bar E_i =  - \bar{f}_{0i}$, the equations of motion for $\phi$ and $\chi$ lead to one unstable mode~\cite{Yamamoto:2022vrh}.

By using the equation of motion for $a$, we have the following conserved charge:
\begin{equation}
Q_a ({\cal C}) = \int_{\cal C} \(- \fr{1}{e^2} * \rd a 
+
\fr{1}{4\pi^2} 
\phi \, 
\rd \chi
\)
= \int_{\cal C} \(- \fr{1}{2 e^2} \epsilon_{\mu\nu\rho} f^{\nu\rho} 
+ 
\fr{1}{4\pi^2} 
\phi \der_\mu \chi
\) {\rd}x^\mu\,.
\end{equation}
In particular, by taking ${\cal C}$ as a line $L$ perpendicular to the electric field, we have 
\begin{equation}
Q (L) 
=
 \int_{L}\(- \fr{1}{e^2} \epsilon_{i 0j } f^{0 j } 
+ \fr{1}{4\pi^2} \phi \der_i \chi\) {\rd}x^i
=
- \fr{1}{e^2}  \int_{L} \bs{E} \times \rd \bs{x}
+ 
\fr{1}{4\pi^2} 
\int_{L} \phi \bs{\na}\chi \cdot \rd\bs{x}\,.
\end{equation}
The first term counts the electric flux $\bs{E}$ through the line $L$. 
The second term
\begin{equation}
 T_{a} (L) = 
\fr{1}{4\pi^2} 
\int_{L} \phi \, 
\rd\chi  
= 
\fr{1}{4\pi^2} 
\int_{L}\phi \bs{\na}\chi \cdot \rd\bs{x}\,,
\end{equation}
which is independent of the metric of the spacetime, can be understood as a generalized magnetic helicity according to the discussion in section~\ref{GMH}. We can see that the initial electric field decreases and that $T_{a}(L)$ increases under the time evolution.

\begin{figure}[t]
\centering  \ig[width=37.5em]{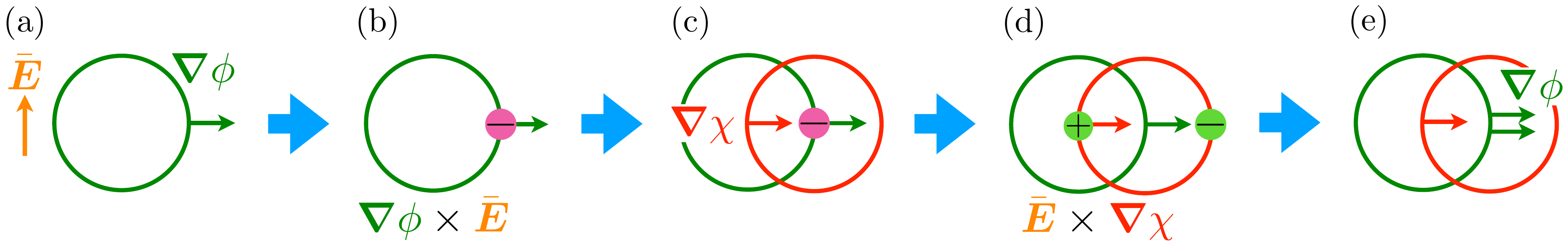} 
\caption{\label{inst3d}Instability in the $(2+1)$-dimensional Goldstone-Maxwell model. The domain walls of $\phi$ and $\chi$ are expressed as the green and red circles, respectively. The minus sign on the pink dot is the sign of the source for $\bs{\na}\chi$, i.e., $\bs{\na} \phi \times \bar{\bs{E}}$.}
\end{figure}

We again give an intuitive understanding of the instability and the growth of $T_{a}(L)$. We begin with the configuration of an infinitesimal domain wall for $\phi$ in the presence of the homogeneous electric field $\bar E_i = - \bar{f}_{0i}$ (figure~\ref{inst3d}(a)). 
Since $\bs{\na}\phi \times \bar{\bs{E}}$ acts as a source for $\bs{\na}\chi$ by the equation of motion for $\chi$, $v^2_\chi \bs{\na}\cdot \bs{\na}\chi = \fr{1}{4\pi^2} \bs{\na}\phi \times \bar{\bs{E}}$ (figure~\ref{inst3d}(b)), a domain wall for $\chi$ is created 
(figure~\ref{inst3d}(c)).
By the equation of motion for $\phi$, $v^2_\phi \bs{\na}\cdot \bs{\na}\phi = \fr{1}{4\pi^2} \bar{\bs{E}} \times \bs{\na}\chi$, on the other hand, $\bar{\bs{E}} \times \bs{\na}\chi$ acts as a source for $\bs{\na}\phi$, which is positive and negative inside and outside of the domain wall, respectively (figure~\ref{inst3d}(d)), leading to the growth of $\bs{\na}\phi$ (figure~\ref{inst3d}(e)).

\begin{figure}[t]
\vspace{1em}
\centering  \ig[height=6em]{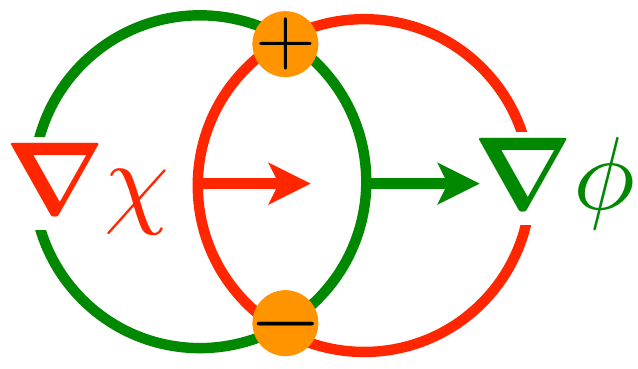} 
\caption{\label{cross} Decrease of the initial electric field. The induced electric charges represented by the orange dots give the electric field whose direction is opposite to that of the initial electric field.}
\end{figure}

\begin{figure}[t]
\begin{center}
 \ig[height=6em]{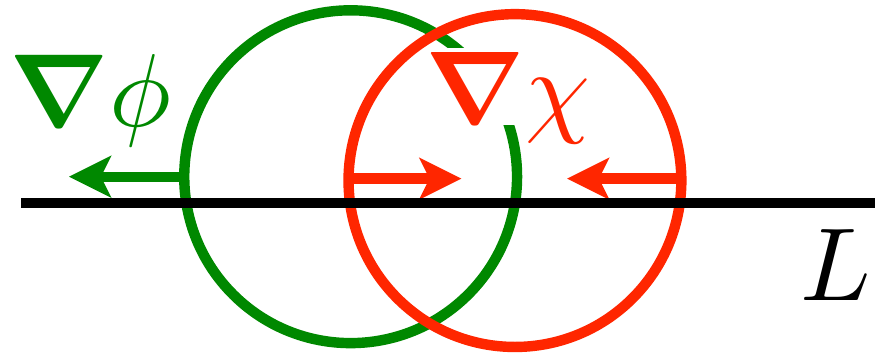} 
\end{center}
\caption{\label{3dlink}
Linking between the domain walls of $\phi$ and $\chi$ on the line $L$.
The black line expresses the line for the integration of the electric flux. The red and green circles are linked on the black line: the intersection of the red circle and black line is surrounded by the intersection of the green circle and black line.}
\end{figure}

Furthermore, we can understand the decrease of the initial electric field and generation of the configuration with the linking number as in figures~\ref{cross} and \ref{3dlink}, respectively. By the equation of motion for the gauge field, $\fr{1}{e^2} \bs{\na}\cdot \bs{E}  = \fr{1}{4\pi^2} \bs{\na}\phi \times \bs{\na}\chi$, electric charges are induced in the region where $\bs{\na}\phi \times \bs{\na}\chi \neq 0$. The induced charges lead to the dielectric polarization, which decreases the electric field inside the domain wall. This configuration has the generalized magnetic helicity related to the linking number. For simplicity, we consider the configuration,
\begin{equation}
    \rd \phi 
    = m_\phi \delta_1 ({\cal S}_\phi ),
    \quad
    \rd \chi 
    = m_\chi \delta_1 ({\cal S}_\chi ),    
\end{equation}
which can be solved as
$  \phi 
    = m_\phi 
    \delta_0 ({\cal V}_{{\cal S}_\phi} ),$
where ${\cal S}_{\phi}$, ${\cal S}_{\chi}$, and ${\cal V}_{{\cal S}_\phi}$ are 2-, 2-, and 3-dimensional subspaces satisfying $\der {\cal V}_{{\cal S}_\phi} = {\cal S}_{\phi}$. In this case, the generalized magnetic helicity $T_a (L)$ can be expressed by the linking number,
\begin{equation}
    T_a (L) 
    =  - \fr{m_\phi m_\chi}{4\pi^2} 
    \int_L \delta_0 ({\cal V}_{{\cal S}_\phi}) 
    \delta_1 ({\cal S}_\chi)
    = -\fr{m_\phi m_\chi }{4\pi^2}
    \link ({\cal S}_\phi , {\cal S}_\chi ; L)\,.
    \end{equation}
The quantity $\link ({\cal S}_\phi, {\cal S}_\chi ; L)$ is the linking number between ${\cal S}_\phi$ and ${\cal S}_\chi$ on $L$ as shown in figure~\ref{3dlink}.

\subsection*{Acknowledgment}
N.~Y.~is supported in part by the Keio Institute of Pure and Applied Sciences (KiPAS) project at Keio University and JSPS KAKENHI Grant Numbers JP19K03852 and JP22H01216. 
R.~Y.~is supported by JSPS KAKENHI Grants Numbers JP21K13928 and JP22KJ3120.

\appendix 

\section{Reality conditions}
\label{realCPI}

In this appendix, we summarize the derivation of the reality condition for the amplitude $A^{\pm}_{\bs{k}, \pm \lambda}$ in section~\ref{CPIreduction}. For simplicity, we focus on the case of the chiral instability, but the following derivation can be extend to the case of generalized chiral instabilities in a straightforward manner.

Since $P_{ij}$ in eq.~(\ref{P_CI}) is a real matrix, the reality condition for $a_j (t, \bs{k})$ amounts to the reality condition for $v_j (t,\bs{k})$, $v_j^* (t,\bs{k}) = v_j (t,-\bs{k})$. On the one hand, the mode expansion of $v_j^* (t, \bs{k})$ is 
\begin{align}
 v_j^* (t,\bs{k}) 
&
 =  
\epsilon^{\rm R}_j
(A^{+* }_{\bs{k},\lambda} {\re}^{-{\ri} t  \sr{|\bs{k}|^2 +\lambda} }
 + A^{-* }_{\bs{k},\lambda}
{\re}^{\ri t \sr{|\bs{k}|^2 +\lambda}})
\nonumber \\
&\quad
+
\epsilon^{\rm L}_j
(A^{+*}_{\bs{k},-\lambda} {\re}^{-{\ri} t  \sr{|\bs{k}|^2 -\lambda} }
 + A^{-*}_{\bs{k},-\lambda}
 {\re}^{{\ri} t \sr{|\bs{k}|^2 -\lambda}})
+\cdots
\end{align}
for $|\bs{k}|^2 > \lambda$, and 
\begin{align}
 v_j^* (t,\bs{k}) 
&
 =
\epsilon^{\rm R}_j
(A^{+*}_{\bs{k},\lambda} {\re}^{-{\ri} t  \sr{|\bs{k}|^2 +\lambda} }
 + A^{-*}_{\bs{k},\lambda} 
{\re}^{{\ri} t \sr{|\bs{k}|^2 +\lambda}})
\\
&\quad
+
\epsilon^{\rm L}_j
(A^{+*}_{\bs{k},-\lambda} {\re}^{ t\sr{\lambda - |\bs{k}|^2} }
 + A^{-*}_{\bs{k},-\lambda}
 {\re}^{- t \sr{\lambda  - |\bs{k}|^2}}
)
+\cdots
\end{align}
for $|\bs{k}|^2 < \lambda$. On the other hand, $v_j(t, -\bs{k})$ is given by the solution to the equation of motion for $\hat{a}_i (t, -\bs{k})$, where the sign of the matrix $\ri k_l \epsilon^{ilj0}$ is flipped, 
\begin{equation}
 (-\der_0^2 - \bs{k}^2) \hat{a}^i (t, -\bs{k})
= -\ri \b\sigma \epsilon^{i l j 0} k_l \hat{a}_j(t,-\bs{k}). 
\end{equation}
By using the orthogonal matrix $P_{ij}$, we can block diagonalize the equation of motion, 
\begin{equation}
 (-\der_0^2 - \bs{k}^2) v_i (t, -\bs{k})
= -\ri \Lambda_{ij} v_j(t,-\bs{k}). 
\end{equation}
and solve the equation in terms of $v_j (t,- \bs{k})$:
\begin{align}
 v_j (t,- \bs{k}) 
&
 = 
\epsilon^{\rm R}_j
(A^{+}_{-\bs{k}, \lambda}  {\re}^{{\ri} t  \sr{|\bs{k}|^2 +\lambda} }
 + A^{-}_{-\bs{k}, \lambda} 
{\re}^{-{\ri} t \sr{|\bs{k}|^2 +\lambda}})
\nonumber \\
&\quad
+
\epsilon^{\rm L}_j
(A^{+}_{-\bs{k}, -\lambda} 
{\re}^{{\ri} t  \sr{|\bs{k}|^2 -\lambda} }
 + A^{-}_{-\bs{k}, -\lambda}
{\re}^{-{\ri} t \sr{|\bs{k}|^2 -\lambda}})
+\cdots
\end{align}
for $|\bs{k}|^2 > \lambda$,
and 
\begin{align}
 v_j (t,-\bs{k}) 
&
 =
\epsilon^{\rm R}_j
(A^{+}_{-\bs{k}, \lambda} {\re}^{{\ri} t  \sr{|\bs{k}|^2 +\lambda} }
 + A^{-}_{\bs{k},\lambda} 
{\re}^{-{\ri} t \sr{|\bs{k}|^2 +\lambda}})
\nonumber \\
&\quad
+
\epsilon^{\rm L}_j
(A^{+}_{-\bs{k}, -\lambda} {\re}^{ t\sr{\lambda - |\bs{k}|^2} }
 + A^{-}_{-\bs{k}, -\lambda} 
 {\re}^{- t \sr{\lambda  - |\bs{k}|^2}}
)
+\cdots
\end{align}
for $|\bs{k}|^2 < \lambda$. Note that the polarization vectors $\epsilon_j^{\rm R, L}$ associated with the negative and positive eigenvalues $\mp \lambda$ are flipped because of the change of the sign in $-\ri \b\sigma \epsilon^{i l j 0} k_l$. Therefore, the reality condition for $A^{\pm}_{\bs{k}, \lambda}$ is 
\begin{equation}
(A^{\pm}_{\bs{k}, \lambda})^* 
= 
A^{\mp}_{-\bs{k}, \lambda}.
\end{equation}
Meanwhile, the reality condition for $A^{\pm}_{\bs{k}, -\lambda}$ is sensitive for the sign of $|\bs{k}|^2 - \lambda$ as
\begin{gather}
(A^{\pm}_{\bs{k}, -\lambda})^* 
= 
A^{\mp}_{-\bs{k}, -\lambda},
 \qtq{for} |\bs{k}|^2 > \lambda,
 \nonumber \\
(A^{\pm}_{\bs{k}, -\lambda})^* 
= 
A^{\pm}_{-\bs{k}, -\lambda} ,
 \qtq{for} |\bs{k}|^2 < \lambda.
\end{gather}

\providecommand{\href}[2]{#2}

\end{document}